\documentclass[lettersize,journal]{IEEEtran}
\usepackage{CustomMacros}
\usepackage{amsmath,amsfonts}
\usepackage{array}
\usepackage[caption=false,font=normalsize,labelfont=sf,textfont=sf]{subfig}
\usepackage{textcomp}
\usepackage{stfloats}
\usepackage{url}
\usepackage{cite}
\usepackage{verbatim}
\usepackage{graphicx}
\usepackage{algorithm}
\usepackage{algpseudocode}
\usepackage{caption}
\usepackage{makecell}

\definecolor{darkblue}{rgb}{0.0, 0.0, 0.55}
\definecolor{cobalt}{rgb}{0.0, 0.28, 0.67}
\definecolor{cardinal}{rgb}{0.77, 0.12, 0.23}
\definecolor{tropicalrainforest}{rgb}{0.0, 0.46, 0.37}
\definecolor{viridian}{rgb}{0.25, 0.51, 0.43}
\definecolor{sapgreen}{rgb}{0.31, 0.49, 0.16}
\definecolor{sacramentostategreen}{rgb}{0.0, 0.34, 0.25}
\definecolor{officegreen}{rgb}{0.0, 0.5, 0.0}

\usepackage{hyperref}
\hypersetup{
    colorlinks,
    linkcolor={darkblue},
    filecolor={darkgreen},
    citecolor={darkblue},
    urlcolor={cobalt}
}

\usepackage{etoolbox}
\newcommand{\zerodisplayskips}{%
  \setlength{\abovedisplayskip}{2.5pt}%
  \setlength{\belowdisplayskip}{2.5pt}%
  \setlength{\abovedisplayshortskip}{2.5pt}%
  \setlength{\belowdisplayshortskip}{2.5pt}}
\appto{\normalsize}{\zerodisplayskips}
\appto{\small}{\zerodisplayskips}
\appto{\footnotesize}{\zerodisplayskips}
\allowdisplaybreaks

\definecolor{azure}{rgb}{0.0, 0.5, 1.0}

\begin{document}

\setlength{\parskip}{0.4em}
\title{Privacy Preserving Semi-Decentralized Mean Estimation over Intermittently-Connected Networks}

\author{Rajarshi Saha, Mohamed Seif, Michal Yemini, Andrea J. Goldsmith, H. Vincent Poor
\thanks{R.~Saha (rajsaha@stanford.edu) (corresponding author) is with the Department of Electrical Engineering, Stanford University, Stanford, CA, USA 94305. (Email: rajsaha@stanford.edu).
M.~Seif (mseif@princeton.edu), A.~J.~Goldsmith (goldsmith@princeton.edu), and H.~V.~Poor (poor@princeton.edu) are with the Department of 
Electrical and Computer Engineering, Princeton University, Princeton, NJ, USA 08540.
M.~Yemini (michal.yemini@biu.ac.il) is with the Department of Engineering, Bar-Ilan University, Ramat-Gan, Israel 5290002.
}
\thanks{
This work was supported by the AFOSR award \#002484665, a Huawei Intelligent Spectrum grant, and NSF grants CCF-1908308 \& CNS-2128448.
}
\thanks{
A preliminary version of this paper was presented at the 2023 IEEE International Symposium on Information Theory, \cite{saha2023isit-p2p-privacy}. 
This work presents a detailed analysis of the algorithm and simulations, along with the derivations of central privacy guarantees, while allowing for heterogeneity in peer-to-peer privacy noise perturbations.
}
}
\date{\today}

\markboth{Manuscript under review}%
{Saha \MakeLowercase{\textit{et al.}}: Privacy Preserving Semi-Decentralized Mean Estimation over Intermittently-Connected Networks}


\maketitle

\begin{abstract}
We consider the problem of privately estimating the mean of vectors distributed across different nodes of an unreliable wireless network, where communications between nodes can fail intermittently.
We adopt a semi-decentralized setup, wherein to mitigate the impact of intermittently connected links, nodes can collaborate with their neighbors to compute a local consensus, which they relay to a central server. 
In such a setting, the communications between any pair of nodes must ensure that the privacy of the nodes is rigorously maintained to prevent unauthorized information leakage. 
We study the tradeoff between collaborative relaying and privacy leakage due to the data sharing among nodes and, subsequently, propose \pricer: {\bf Pri}vate {\bf C}ollaborative {\bf E}stimation via {\bf R}elaying -- a differentially private collaborative algorithm for mean estimation to optimize this tradeoff. 
The privacy guarantees of \pricer~ arise $\rm \bf (i)$ \textit{\bf implicitly}, by exploiting the inherent stochasticity of the flaky network connections, and $\rm \bf (ii)$ \textit{\bf explicitly}, by adding Gaussian perturbations to the estimates exchanged by the nodes.
Local and central privacy guarantees are provided against eavesdroppers who can observe different signals, such as the communications amongst nodes during local consensus and (possibly multiple) transmissions from the relays to the central server.
We substantiate our theoretical findings with numerical simulations.
Our implementation is available at \href{https://github.com/rajarshisaha95/private-collaborative-relaying}{https://github.com/rajarshisaha95/private-collaborative-relaying}.
\end{abstract}


\section{Introduction}
\label{sec:introduction}
\IEEEPARstart{D}{istributed} Mean Estimation (DME) is a fundamental statistical problem that arises in several applications, such as model aggregation in federated learning (FL) \cite{mcmahan2017communication}, distributed K-means clustering \cite{balcan2013distributed}, and distributed power iteration \cite{suresh2017distributed}.
DME presents several practical challenges, which prior studies \cite{kar2008distributed,tandon2017gradient, kar_2008_topology_design, kar_2012_nonlinear_observation_model, jhunjhunwala2021leveraging} have considered, including the problem of straggler nodes, where distributed nodes cannot send their data to the aggregator, or parameter server (PS).
In practical systems, typically, there are two types of stragglers: $\rm (i)$ {\it computation stragglers}, which cannot finish their local computation within a deadline, and $\rm (ii)$ {\it communication stragglers}, in which nodes cannot transmit their updates due to communication blockage \cite{gapeyenko_mobile_blockers, yasamin, pappas, zavlanos, yemini_et_al:globecom2020, yemini_cloud_cluster}. 
For instance, nodes may become communication stragglers when their wireless communication channel to the PS is temporarily blocked.
Unlike computation stragglers, the problem of communication stragglers can be mitigated by relaying the updates/data via neighboring nodes that have better connectivity to the PS.

The problem of intermittent node participation, primarily in the context of Federated Learning, has been studied in prior works such as \cite{gu2021fast, Yan2020DistributedNO, wang_2022, pmlr-v130-ruan21a, anarchic_FL, Sun_2024, wang2024a}.
The collaborative relaying approach for mitigating node intermittency was proposed in \cite{saha2022colrel,yemini2022_ISIT_robust,yemini2022robust_TWC}, where it was shown that in the presence of communication stragglers, collaborative relaying can improve the accuracy of mean estimation and also enhance the convergence rate of FL algorithms.
While these works show the benefit of relaying, they do not secure the client data against the potential breach of privacy due to the additional exchange of information across the nodes.
To mitigate the privacy leakage in DME via collaborative relaying, we must define the notions of peer-to-peer privacy, and {\it central} privacy at the relay.
Within the context of distributed learning, \textit{local differential privacy} (LDP) \cite{dwork2014algorithmic} has been adopted as a standard notion of privacy, in which a node perturbs and discloses a sanitized version of its data to an \textit{untrusted} server. 
LDP ensures that the statistics of the node's output observed by eavesdroppers are indistinguishable regardless of the realization of any input data. 
In this paper, we focus on the \textit{node-level} LDP where neighboring nodes, as well as any eavesdropper that can observe the local node-node transmissions during collaborations, cannot infer the realization of a node's data.

There has been extensive research into the design of distributed learning algorithms that are both communication efficient and private (see \cite{kairouz2021advances} for a comprehensive survey and references therein). 
It is worth noting that LDP requires a significant amount of perturbation noise to ensure reasonable privacy guarantees. 
Nonetheless, the amount of perturbation noise can be significantly reduced by considering the intermittent connectivity of nodes in the learning process \cite{balle2018privacy}. 
The intermittent connectivity in DME amplifies the privacy guarantees in the sense that it provides an increased level of anonymity due to partial communication with the server. 
Various random node participation schemes such as Poisson sampling \cite{zhu2019poission}, importance sampling \cite{luo2022tackling, rizk2022federated}, and sampling with/without replacement \cite{balle2018privacy}, have been proposed to further improve the utility-privacy tradeoff in distributed learning.
In addition, \cite{balle2020privacy} investigated the privacy amplification in federated learning via random check-ins and showed that the privacy leakage scales as $O(1/\sqrt{n})$, where $n$ is the number of nodes. 
Random node participation reduces the amount of noise required to achieve the same levels of privacy that are achieved without sampling.

Existing works in the privacy literature, such as \cite{dwork2014algorithmic,kairouz2021advances,balle2018privacy,zhu2019poission,balle2020privacy, asi_2022b, cummings_2022, isik2023exact, gaboardi19a, Xue2021MeanEO}, do not consider intermittent connectivity along with collaborative relaying, where nodes share their local updates to mitigate the randomness in network connectivity \cite{yemini2022_ISIT_robust,yemini2022robust_TWC,saha2022colrel}.
In this work, we propose \pricer~ ({\bf Pri}vate {\bf C}ollaborative {\bf E}stimation via {\bf R}elaying) for DME under an intermittent connectivity assumption, and study its tradeoff between collaborative relaying and privacy leakage due to data sharing among nodes.
\pricer~ is a two-stage algorithm, in which nodes improve their connectivity to the PS by sharing their weighted and privatized data with neighbors.
For a given network topology, \pricer~ finds the best collaboration scheme by solving a joint weight and privacy-noise variance optimization problem, subject to a set of peer-to-peer privacy constraints.
Privacy constraints, which are a function of how much nodes trust each other, manifest as cone constraints, and we iteratively obtain a close-to-optimal solution using projected gradient descent.
Moreover, for a symmetric network topology, we derive the optimal solution in closed form.

In a distinct line of work, {\it over-the-air computation} (OAC) strategies have been proposed, which exploit the signal superposition property of non-orthogonal links between nodes and the PS, for mean estimation \cite{oac1, oac2, oac3}.
OAC aggregation schemes provide a certain degree of anonymity to the participating nodes, consequently enhancing their privacy \cite{oac_privacy1, oac_privacy2}.
\pricer~ is compatible with OAC aggregation schemes and so, we also study the privacy leakage from \pricer~ at any aggregating node acting as a relay.
Subsequently, we provide the privacy guarantees for protecting the anonymity and local data of any node which collaborates with the relay.

\subsection{Main Contributions}
\label{main_contributions}

Our main contributions can be summarized as follows:
\begin{itemize}
    \item We generalize the collaborative relaying approach proposed in \cite{saha2022colrel,yemini2022_ISIT_robust,yemini2022robust_TWC} to jointly consider the implications of privacy leakage due to collaboration.
    In our proposed framework, nodes improve their connectivity to the PS by exchanging scaled and perturbed versions of their local private data with their neighbors, while taking into account the trustworthiness of their neighbors.
    \item We introduce the individualized notion of local peer-to-peer privacy constraints as a measure to quantify the trustworthiness between any pair of nodes.
    \item Our algorithm optimizes private collaborative relaying so as to minimize the MSE and to control the bias accumulated at the PS with a bias-regularization term.
    \item We study the central privacy leakage at the relays and the PS due to the redundancy introduced from collaboration.
    Specifically, we consider the privacy guarantees for preserving the anonymity as well as the local data of any node participating in the collaboration process.
    We study our framework in detail for the special case of an Erdös-Rényi network topology.
    \item We validate our framework via numerical simulations and demonstrate its efficacy in overcoming intermittent network connectivity in the presence of privacy constraints for mean estimation and $\rm K$-means clustering tasks.
\end{itemize}

\subsection{Paper Organization}
\label{subsec:paper_organization}

The remainder of the paper is as follows.
In \S \ref{sec:system-model-for-private-collaboration}, we introduce our system model and proposed algorithm, followed by mean-squared error (MSE) and privacy analyses in \S\ref{sec:estimation-error-analysis} and \S \ref{sec:privacy-analysis}, respectively.
We propose our privacy-constrained joint weight and variance optimization in \S \ref{sec:privacy-constrained-weight-optimization}.
Finally, we demonstrate the efficacy of \pricer~ via numerical simulations for different datasets and network topologies in \S \ref{sec:numerical-simulations}, and conclude in \S \ref{sec:conclusions}.
Detailed derivations are provided in the appendices.

\section{System model for private collaboration}
\label{sec:system-model-for-private-collaboration}

\begin{figure}[htbp]
  \centering
  \includegraphics[width=0.4\textwidth]{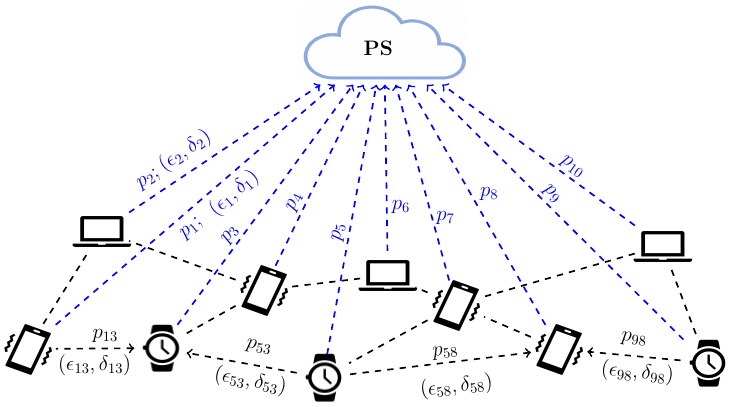}
  \captionsetup{font=small}
  \caption{Intermittently connected network. Dotted lines indicate intermittent node-PS and node-node connections. Communication between any two nodes must satisfy DP constraints.}
  \label{fig:system-model-pricer}
  \vspace{-5mm}
\end{figure}

Consider a distributed system with $n$ nodes, where node $i$ has an observation vector $\xv_i \in \Real^d$, $\lVert \xv_i \rVert \leq {\rm R}$ for some known ${\rm R} > 0$. 
The nodes communicate with a PS and with each other over intermittent links with the goal of estimating the mean, $\xvo \triangleq \frac{1}{n}\sum_{i=1}^{n}\xv_i$ at the PS as shown in Fig. \ref{fig:system-model-pricer}, so as to minimize the MSE, given by $\Ecal \triangleq \mathbb{E}\lVert\xvoh - \xvo\rVert^2$.

\subsection{Communication model}
\label{subsec:communication-model}

We consider a setup where communication links are either unavailable or available perfectly, i.e., they do not suffer from any channel imperfections. 
Such links are modeled using $0/1$ Bernoulli random variables.
As shown in Fig. \ref{fig:system-model-pricer}, node $i$ can communicate with the PS with probability $p_i$, with the link modeled using a Bernoulli random variable $\tau_i \sim {\rm Ber}(p_i)$.
$\tau_i = 0$ implies a communication blockage between node $i$ and the PS, whereas $\tau_i = 1$ implies the presence of a communication opportunity. 
Similarly, node $i$ can communicate with another node $j$ with probability $p_{ij}$, i.e., $\tau_{ij} \sim {\rm Ber}(p_{ij})$.
The links between different node pairs are assumed to be statistically independent, i.e., $\tau_i \perp \tau_j$ for $i \neq j$, $\tau_{ij} \perp \tau_{ml}$ for $(i,j) \neq (m,l)$, $(j,i) \neq (m,l)$, and $\tau_{ij} \perp \tau_l$ for $i,j,l$.
The correlation of channels between a pair of nodes $i,j$ is denoted by $\Em_{\{i,j\}} \equiv \mathbb{E}[\tau_{ij}\tau_{ji}]$.
We assume that $\mathrm{E}_{\{i,j\}}\geq p_{ij}p_{ji}$, i.e., $\mathbb{P}(\tau_{ij}=1|\tau_{ji}=1)\geq \mathbb{P}(\tau_{ij}=1)$.
Furthermore, $p_{ii} = 1 \; \forall \; i \in [n]$, and if node $i$ can never transmit to $j$, we have $p_{ij} = 0$. 
We denote $\pv \equiv (p_1, \ldots, p_n)$ and $\Pv \equiv (p_{ij})_{i,j \in [n]} \in [0,1]^{n \times n}$.

\subsection{Privacy model}
\label{subsec:privacy-model}

The nodes are assumed to be {\it honest but curious}. 
They are {\it honest} because they faithfully perform their role in the system, i.e., compute the aggregate of the received signals.
However, they are {\it curious} as they might be interested in learning sensitive information about other nodes.
Each node uses an independent local additive noise mechanism to ensure the privacy of its transmissions to neighboring nodes.
Since there are several transmissions across nodes, and between nodes and the PS, there can be different sources of privacy leakage depending on what the eavesdropper (which can be a node or an external agent) has access to. 
We consider local privacy constraints, wherein node $i$ trusts another node $j$ to a certain extent, and hence, randomizes its own data accordingly when sharing with node $j$ using a synthetic Gaussian noise to respect the privacy constraint of the link $i \to j$.
In addition to this, we also consider central privacy at each node when it acts as a relay and does local aggregation in Alg. \ref{algo:pricer-local-aggregation}, as well as the central privacy at the PS (Alg. \ref{algo:pricer-global-aggregation}).
A brief refresher on differential privacy and Gaussian mechanism is presented in App. \ref{app:preliminaries}. 
We formally analyze the privacy guarantees in \S \ref{sec:privacy-analysis}.

\subsection{Private collaborative relaying for mean estimation}
\label{subsec:private-collaborative-relaying-for-mean-estimation}

We now introduce \pricer, which is a two-stage semi-decentralized algorithm for estimating the mean.
In the first stage, each node $j$ sends a scaled and privacy noise-added version of its data to a neighboring node $i$ over the intermittent link $\tau_{ji}$.
The transmitted signal from node $j \to i$ is
\begin{equation}
    \xtv_{ji} = \tau_{ji}(\alpha_{ji}\xv_j + \nv_{ji}).
\end{equation}
Here, $\alpha_{ji} \geq 0$ is the weight used by node $j$ while transmitting to node $i$, and $\nv_{ji} \sim \Ncal(\mathbf{0}, \sigma_{ji}^2\Iv_d)$ is the multivariate Gaussian noise added by node $j$.
We denote the weight matrix by $\Av \equiv (\alpha_{ij})_{i, j \in [n]}$, and the noise variance matrix by $\Sigmav = (\sigma_{ij})_{i,j \in [n]}$.
Consequently, node $i$ computes the local aggregate of all received signals as
\begin{equation}
    \xtv_i = \sum_{j \in [n]}\tau_{ji}(\alpha_{ji}\xv_j + \nv_{ji}).
\end{equation}

By observing $\xtv_{ji}$ or $\xtv_i$, an eavesdropper (possibly including node $i$) should not be able to distinguish between the events when node $j$ contains the data $\xv_j$ versus when it contains some other data $\xv_j'$.
In addition to this, by observing $\xtv_i$, an eavesdropper should also not be able to tell when node $j$ participated in the aggregation at node $i$ or not.
We assume that the privacy noise added by different nodes are uncorrelated, i.e., $\mathbb{E}[\nv_{il}^\top\nv_{jm}] = 0$ for all $i,j,l,m$ as long as $i,j,l,m$ are not all equal. 
In the second stage, each node $i$ transmits $\xtv_i$ to the PS over the intermittent link $\tau_i$, and the PS computes the global estimate.
The Pseudocode for \pricer~ is given in Algs. \ref{algo:pricer-local-aggregation} and \ref{algo:pricer-global-aggregation}.

\begin{algorithm}[H]
    \caption{{\bf \pricer}-Stage 1: Local aggregation}
    \label{algo:pricer-local-aggregation}
    \begin{algorithmic}
    \State {\bf Input}: Non-negative weight matrix $\Av$
    \State {\bf Output}: $\widetilde{\xv}_i$ for all $i\in[n]$
    \For{$i=1$ to $n$}
    \State Locally generate $\xv_i$
    \State Transmit $\xtv_{ij} = \alpha_{ij}\xv_i + \nv_{ij}$ to nodes $j \neq i$
    \State Receive $\xtv_{ji} = \tau_{ji}(\alpha_{ji}\xv_j + \nv_{ji})$ from $j \neq i$
    \State Set $\xtv_{ii} = \alpha_{ii}\xv_i + \nv_{ii}$
    \State Locally aggregate: $\xtv_i=\sum_{j\in[n]}\xtv_{ji}$
    \State Transmit $\xtv_i$ to the PS
    \EndFor
    \end{algorithmic}
\end{algorithm}

\begin{algorithm}[H]
    \caption{{\bf \pricer}-Stage 2: Global aggregation}
    \label{algo:pricer-global-aggregation}
    \begin{algorithmic}
    \State {\bf Input}: $\tau_i\xtv_i$ for all $i \in[n]$
    \State {\bf Output}: Estimate of the mean at the PS: $\xvoh$
    \For{$i=1$ to $n$}
    \State Receive $\tau_i\xtv_i$
    \EndFor
    \State Aggregate the received signals: $\xvoh = \frac{1}{n}\sum_{i \in [n]}\tau_i\xtv_i$
    \end{algorithmic}
\end{algorithm}

\section{Estimation error analysis}
\label{sec:estimation-error-analysis}

The goal of \pricer~ is to obtain an estimate of $\xvo$ at the PS.
Since each node sends its data to all other neighboring nodes, the PS receives multiple copies of the same data, which can introduce some bias.
Let us denote $S_i \triangleq \sum_{j \in [n]}p_jp_{ij}\alpha_{ij}$,
i.e., after relaying from all the nodes, the PS receives an expected contribution of node $i$ as $S_i\xv_i$, implying an expected bias of $(S_i - 1)\xv_i$.
Let us also define the Euclidean ball, $\mathbb{B}_d(\Rm) = \{\xv \in \Real^d \; \norm{\xv} \leq \Rm\}$.
The following result upper bounds the MSE of \pricer.

\begin{theorem}
\label{thm:pricer-estimation-error-analysis}
Suppose for some $\Rm > 0$, $\xv_i \in \mathbb{B}_d(\Rm)$ for all $i \in [n]$.
Given $\pv, \Pv$, $\Av$, and $\Sigmav$, the MSE of \pricer~ is upper bounded as:
\begin{equation}
    \label{eq:pricer-error-upper-bound}
    \sup_{\{\xv_i \in \mathbb{B}_d(\Rm)\}}\Ebb\lVert\xvoh - \xv\rVert^2 \leq \sigma_{\rm tiv}^2(\pv,\Pv, \Av) + \sigma_{\rm pr}^2(\pv, \Pv, \Sigmav),
\end{equation}
where the expectation is over the stochasticity of intermittent connections and the randomness due to privacy noise.
Here, $\sigma_{\rm tiv}^2(\pv, \Pv, \Av)$ is the {\bf topology-induced variance} introduced by the stochasticity due to intermittent topology given by

\begin{small}
\begin{align}
\label{eq:topology-induced-variance-expression}
    &\sigma_{\rm tiv}^2(\pv, \Pv, \Av) \triangleq \frac{{\rm R}^2}{n^2}\left[\sum_{i,j \in [n]}p_jp_{ij}\Paren{1 - p_{ij}}\alpha_{ij}^2\right. \nonumber\\
    &\hspace{30mm}+\left.\sum_{i,j,l \in [n]}p_j(1 - p_j)p_{ij}p_{lj}\alpha_{ij}\alpha_{lj}\right. \nonumber \\
    &\left.+ \sum_{i,j \in [n]}p_ip_j(\Em_{\{i,j\}} - p_{ij}p_{ji})\alpha_{ij}\alpha_{ji} + \left(\sum_{i \in [n]}(S_i - 1)\right)^2\right],
\end{align}
\end{small}
and $\sigma_{\rm pr}^2(\pv, \Pv, \Sigmav)$ is the {\bf privacy-induced variance} given by
\begin{equation}
\label{eq:privacy-induced-variance-expression}
    \sigma_{\rm pr}^2(\pv, \Pv, \Sigmav) \triangleq \frac{d}{n^2}\sum_{i,j \in [n]}p_jp_{ij}\sigma_{ij}^2.
\end{equation}
\end{theorem}

Thm. \ref{thm:pricer-estimation-error-analysis} is derived in App. \ref{app:pricer-estimation-error-analysis}.
From \eqref{eq:pricer-error-upper-bound}, we see that $\sigma_{\rm pr}^2(\pv,\Pv, \Sigmav)$ is the price of privacy.
For a non-private setting, i.e., $\sigma_{ij} = 0 \; \forall \; i,j$, the privacy induced variance $\sigma_{\rm pr}^2(\pv, \Pv, \Sigmav) = 0$, and Thm. \ref{thm:pricer-estimation-error-analysis} simplifies to \cite[Thm. 3.2]{saha2022colrel}.
In the following section, we present privacy guarantees leading to a choice of weight matrix $\Av$ for the optimal utility (MSE) -- privacy tradeoff for \pricer.

\section{Privacy analysis}
\label{sec:privacy-analysis}

In Fig. \ref{fig:system-model-pricer}, there are several transmissions across nodes, and between nodes and the PS.
Consequently, there can be different sources of privacy leakage depending on what the eavesdropper can access. 
The privacy guarantees of \pricer~ result from two effects: $\rm (i)$ the local noise added at each node, and $\rm (ii)$ the intermittent nature of the connections.
We consider guarantees for protecting the local data of each node, and also for protecting the identity of a particular node in the local aggregation steps.
We formalize these guarantees below.

\subsection{Local privacy guarantee for node-node communications}
\label{subsec:local-privacy-node-node}
 
We first consider the case in which an eavesdropper can observe the transmission from node $i$ to node $j$ during the local-aggregation step of \pricer.
Let us denote the dataset of node $i$ as $\Dcal_i$.
For DME, $\Dcal_i$ is a singleton set and by observing the transmission from node $i$ to node $j$, the eavesdropper (which can possibly include node $j$ itself) should not be able to differentiate between the events $\xv_i \in \Dcal_i$ and $\xv_i' \in \Dcal_i$, where $\xv_i' \neq \xv_i$, formalized in \eqref{eq:node-node-privacy-definition}.
The following result, which is proved in App. \ref{app:proof-local-privacy-node-node}, states our local privacy guarantee.

\begin{theorem}
\label{thm:local-privacy-node-node}
Given $\nv_{ij}\sim \Ncal(\mathbf{0}, \sigma_{ij}^2\Iv_d)$, $\xv_i, \xv_i' \in \mathbb{B}_d(\Rm)$, and any $\delta_{ij} \in (0,1]$, for $(\epsilon_{ij},p_{ij}\delta_{ij})$ that satisfy
\begin{equation}
\label{eq:eps_expression_node_node_privacy}
    \epsilon_{ij} =
    \begin{cases}
    \Br{2\log\Paren{\frac{1.25 }{\delta_{ij}}}}^{\frac{1}{2}}\frac{2\alpha_{ij}{\rm R}}{\sigma_{ij}} \; \; \;  \text{ if } \; p_{ij} > 0, \;\; \text{ and, } \\
    \hspace{15mm}0 \; \;\hspace{16.5mm} \text{ if }\;   p_{ij} = 0,
    \end{cases}
\end{equation}
the transmitted signal from node $i$ to node $j$, i.e., $\xtv_{ij} = \tau_{ij}(\alpha_{ij}\xv_i +\nv_{ij})$ is $(\epsilon_{ij}, p_{ij}\delta_{ij})$-differentially private.
In other words, for any measurable set $\Scal$, $\xtv_{ij}$ satisfies
\begin{equation}
\label{eq:node-node-privacy-definition}
    \Pr\Paren{\xtv_{ij} \in \Scal\mid \xv_i \in \Dcal_i} \leq e^{\epsilon_{ij}}\Pr\Paren{\xtv_{ij} \in \Scal \mid \xv_i' \in \Dcal_i} + p_{ij}\delta_{ij}.
\end{equation}
\end{theorem}

Here, $\epsilon_{ij}$ is a measure of how much node $i$ trusts node $j$; a smaller value implying less trust, and hence requirement of a stricter privacy guarantee.

\begin{figure*}[b]
\hrulefill 
\vspace*{4pt} 
\begin{align}
\label{eq:choice_of_t_bernstein_inequality}
    r = \frac{{\max_{k \in [n]\setminus\{j\}} \sigma_{kj}^{2}}/{3} + \sqrt{{\max_{k \in [n]\setminus\{j\}} \sigma_{kj}^{2}}/{9}  + 4 (\sum_{k \in [n]\setminus\{j\}} p_{kj} (1-p_{kj}) \sigma_{kj}^{4})/\log(2/\delta')}}{2/\log(2/\delta')}
\end{align}
\vspace*{4pt} 
\end{figure*}

\subsection{Central privacy guarantees at a relay}
\label{subsec:central-privacy-guarantee-at-relay}

After each node $j$ receives signals from its neighbors, it locally aggregates them, which is analytically captured as, $\xtvu_j = \sum_{k \in [n]\setminus\{j\}}\tau_{kj}(\alpha_{kj}\xv_k + \nv_{kj})$.
It then acts as a {\bf relay} to transmit $\xtv_j = \xtvu_j + \xtv_{jj}$ to the PS.
Here, node $j$ sets $\xtv_{jj} \triangleq \alpha_{jj}\xv_{j} + \nv_{jj}$ as in Alg. \ref{algo:pricer-local-aggregation}.
We consider the {\it central} privacy leakage of relay $j$ when an eavesdropper (possibly node $j$ itself) can observe $\xtvu_j$.
Privacy guarantees are a consequence of $\rm (i)$ the random participation of nodes while computing $\xtvu_j$ at relay $j$, and $\rm (ii)$ the aggregated noise in $\xtvu_j$ as a consequence of local Gaussian perturbations added by the participating nodes. 

Let $\Rcal_j \triangleq \{k \neq j \mid \tau_{kj} = 1\} \subseteq [n]$ denote the random set of nodes participating at relay $k$.
Note that the effective noise variance at relay $j$, denoted as $\zeta_{j} \triangleq \sum_{k \in [n]\setminus\{j\}} \tau_{kj} \sigma_{kj}^{2} = \sum_{k \in \Rcal_j}\sigma_{kj}^2$, is a function of $\Rcal_j$, excluding relay $j$.
To account for this stochasticity, we condition on the event when the deviation of this aggregated noise $\zeta_{j}$ from its mean, i.e., $\zetao_{j} = \sum_{k \in [n]\setminus\{j\}} p_{kj} \sigma_{kj}^{2}$, is small.
Our central privacy guarantees are formally stated in Thms. \ref{thm:central_privacy_node_identity_protection} and \ref{thm:central_privacy_node_dataset_protection}

{\bf Privacy guarantee for protecting the identity of any node}: For any node $j$ acting as a relay, only a random subset of its neighbors participate in computing $\xtvu_j$.
We are interested in protecting the identity of participating nodes.
Specifically, by observing $\xtvu_j$, an eavesdropper (possibly including node $j$) should not be able to confidently infer whether a particular node $i$ communicated with node $j$. 
Suppose the local aggregation at node $j$ takes place blindly, i.e., it honestly aggregates all the signals it receives anonymously.
This would happen, for example, in over-the-air schemes, wherein different nodes simultaneously send their data to an aggregating node over a shared non-orthogonal wireless medium.
In this setting, aggregation at the PS happens because the wireless channel combines all signals ``over-the-air''.
In such a situation, the aggregating node $j$ is assumed to be \textit{honest-but-curious}, and it should {\bf not} be able to infer the identity of a specific participating node.

Let $\xtvu_j^{(-i)} \triangleq \sum_{k \in \Rcal_j\setminus\{i\}}\Paren{\alpha_{kj}\xv_k + \nv_{kj}}$ be the aggregate at the $j^{\rm th}$ node when the contribution of node $i$ is removed, in case it participated.
Thm. \ref{thm:central_privacy_node_identity_protection}, which is proved in App. \ref{app:privacy_node_identity_at_relay}, formally states this guarantee wherein by observing $\xtvu_j$, an eavesdropper would not be able to distinguish between the presence or absence of a specific node $i$ amongst the set of nodes that participate in the local aggregation at node $j$.

\begin{theorem}\label{thm:central_privacy_node_identity_protection}
{\bf [Protecting node identity]}
Given privacy noises $\nv_{kj} \sim \Ncal(\mathbf{0}, \sigma_{kj}^2\Iv_d)$, $\zeta_{j} = \sum_{k \in [n]\setminus\{j\}} \tau_{kj} \sigma_{kj}^{2}$ with $\tau_{kj} \sim {\rm Ber}(p_{jk})$, $\zetao_j = \sum_{k \in [n]\setminus\{j\}}p_{kj} \sigma_{kj}^{2}$, $r > 0$, and $\delta' \in (0,1)$ such that $\Pr(|\zeta_{j} - \zetao_j| \geq r) \leq \delta'$, for pairs $(\epsilon_{ij}^{(p)}, p_{ij}(\delta_{j}^{(p)} + \delta'))$ satisfying
\begin{equation}
    \epsilon_{ij}^{(p)} = 
    \begin{cases}
        {\Br{2\log\Paren{\frac{1.25}{\delta_{j}^{(p)}}}}^{\frac{1}{2}}\frac{\alpha_{ij}{\rm R}}{\sqrt{\zetao_j - r}}}  \; \; \; \text{ if } \; p_{ij} > 0, \;\; \text{ and, } \\
        \hspace{19mm}0 \; \;\hspace{16mm}  \text{ if } \;\; p_{ij} = 0,
    \end{cases}
\end{equation}
the aggregated signal at node $j$, i.e., $\xtvu_j = \sum_{k \in \Rcal_j}(\alpha_{kj}\xv_k + \nv_{kj})$ is $(\epsilon_{ij}^{(p)}, p_{ij}(\delta_{j}^{(p)} + \delta'))$ differentially private with respect to protecting the identity of any participating node $i$.
In other words, for any measurable set $\Scal$,
\begin{equation}
\label{eq:def_node_identity_protection_relay}
   \Pr\left(\xtvu_j^{(-i)} \in \Scal\right) \leq e^{\epsilon_{ij}^{(p)}}\Pr\left(\xtvu_j \in \Scal\right) + p_{ij}(\delta_{j}^{(p)} + \delta').
\end{equation}
\end{theorem}

{\bf Privacy guarantee for protecting the local data of any node}:
We also consider the privacy guarantee for protecting the local data of any node $i$ when the eavesdropper can observe $\xtvu_j$.
This privacy guarantee for protecting the local data of node $i$ implies that if $\xv_i = \uv$ is perturbed to be $\xv_i = \uv'$, the signal aggregated at node $j$ to the PS, i.e., $\xtvu_j$, should not change significantly.  
The following result (derived in App. \ref{app:privacy_local_data_at_relay}) formalizes this.

\begin{theorem}
\label{thm:central_privacy_node_dataset_protection}
{\bf [Protecting local data at nodes]}
Given privacy noises $\nv_{kj} \sim \Ncal(\mathbf{0}, \sigma_{kj}^2\Iv_d)$, $\zeta_{j} = \sum_{k \in \Rcal_j}\sigma_{kj}^{2}$, $\zetao_j = \sum_{k \in [n]\setminus\{j\}}p_{kj} \sigma_{kj}^{2}$, $r > 0$, and $\delta' \in (0,1)$ such that $\Pr(|\zeta_{j} - \zetao_j| \geq r) \leq \delta'$, for $(\epsilon_{ij}^{(d)},p_{ij} (\delta_{j}^{(d)} + \delta'))$ satisfying
\begin{equation}
    \epsilon_{ij}^{(d)} = 
    \begin{cases}
        2 \cdot {\Br{2\log\Paren{\frac{1.25}{\delta_{j}^{(d)}}}}^{\frac{1}{2}}\frac{\alpha_{ij}{\rm R}}{\sqrt{\zetao_j - r}}}  \; \; \; \text{ if } \; p_{ij} > 0, \;\; \text{ and, } \\
        \hspace{19.5mm}0 \; \;\hspace{18.5mm} \; \text{ if } \;\; p_{ij} = 0,
    \end{cases}
\end{equation}
the aggregate at node $j$, i.e., $\xtvu_j = \sum_{k \in \Rcal_j}(\alpha_{kj}\xv_k + \nv_{kj})$ is $(\epsilon_{ij}^{(d)}, p_{ij}(\delta_{j}^{(d)} + \delta'))$ differentially private with respect to protecting the local data of any node $i$.
In other words, for any $\uv, \uv' \in \mathbb{B}_d(\Rm)$ and measurable set $\Scal$, we have
\begin{align}
\label{eq:def_node_data_protection_relay}
    \Pr\Paren{\xtvu_j \in \Scal \mid \xv_i = \uv} &\leq e^{\epsilon^{(d)}_{ij}} \Pr\Paren{\xtvu_j \in \Scal \mid \xv_i = \uv'} \nonumber\\
     &\qquad\qquad + p_{ij}(\delta_{j}^{(d)} + \delta').
\end{align}
\end{theorem}

In both Thms. \ref{thm:central_privacy_node_dataset_protection} and \ref{thm:central_privacy_node_identity_protection}, we can obtain the parameters $(r,\delta')$ using Bernstein's inequality for $\zeta_j = \sum_{ k \in \Rcal_j}\sigma_{kj}^2$. 
The following corollary, derived in App. \ref{subapp:bernstein_inequality}, gives the smallest value for $r$ that satisfies $\Pr(|\zeta_{j} - \zetao_j| \geq r) \leq \delta'$.
\footnote{One also can use Hoeffding's inequality for bounding this probability. The main drawback of this technique is that it does not capture the connectivity probabilities $p_{ij}$'s. 
Moreover, it can be readily verified that Bernstein's inequality is tighter, specially when $p_{ij} \ll 1$.}  

\begin{corollary} 
\label{cor:choice_of_t_bernstein_inequality}
{\bf [Bernstein parameter]} For a given $\delta' \in (0,1]$, choosing $r$ as in \eqref{eq:choice_of_t_bernstein_inequality} ensures ${\Pr(|\zeta_{j} - \zetao_j| \geq r) \leq \delta'}$.
\end{corollary}

\subsection{Central privacy at parameter server}
\label{subsec:central-privacy-multiple-relays}

We now consider the central privacy guarantees at the PS which can observe the outputs of multiple nodes acting as relays. 
Since the PS receives multiple copies of the data of any specific client $i$ (upto scaling and additive noise) from multiple node relays, the privacy guarantee at the PS can be obtained using composition theorem \cite{kairouz2015composition}.
Let $\yv_j = \tau_j\xtv_j$, where $\tau_j \sim {\rm Ber}(p_j)$, denote the signal received by the PS from the $j^{\rm th}$ relay.
We consider protecting the identity and the local data of a particular node $i$.
In Thm. \ref{thm:central_privacy_at_PS} below (proved in App. \ref{app:proof_central_privacy_at_PS}), we formalize this guarantee by considering that the PS can observe outputs from each of the node relays, i.e., $\{\yv_j\}_{j \in [n]}$.

\begin{theorem}
\label{thm:central_privacy_at_PS}
{\bf [Central privacy at PS]}
For all ${j \in [n]}$, consider we have privacy noises ${\left\{\nv_{kj} \sim \Ncal(\mathbf{0}, \sigma_{kj}^2\Iv_d)\right\}_{k \in [n]}}$, ${\zeta_{j} = \sum_{k \in \Rcal_j} \sigma_{kj}^{2}}$, $\zetao_j = \sum_{k \in [n]\setminus\{j\}}p_{kj} \sigma_{kj}^{2}$, $r_j > 0$, and $\delta_j' \in (0,1)$ such that $\Pr(|\zeta_{j} - \zetao_j| \geq r_j)\leq \delta_j'$.
Furthermore, let $\yv_j = \tau_j\xtv_j$, where $\tau_j \sim {\rm Ber}(p_j)$, and $\yv_j^{(-i)} = \tau_j\xtv_j^{(-i)}$, where $\xtv_j^{(-i)} = \sum_{k \in \Rcal_j \cup \{j\} \setminus\{i\}}\Paren{\alpha_{kj}\xv_k + \nv_{kj}}$.
Suppose the PS observes $(\yv_1, \ldots, \yv_n)$ from all the nodes acting as relays.
Then, for a specific node $i$, for any measurable set $\Scal^n \subseteq \Real^{dn}$,
\begin{equation}
\label{eq:central_privacy_PS_node_identity}
    \Pr(\{\yv_j\}_{j \in [n]} \in \Scal^n) \leq e^{\epsilont_p}\Pr(\{\yv_j^{(-i)}\}_{j \in [n]} \in \Scal^n) + \deltat_p,
\end{equation}
where,
\[{\epsilont_p = \sum_{\substack{j \in [n],\\ p_{ij} > 0}}\epsilont_{ij}^{(p)}},\quad {\deltat_p = \delta\sum_{j \in [n]}p_j},\quad {\delta \in \left(0, {\rm min}_{\substack{j \in [n],\\ p_{ij} > 0}}\{p_{ij}\}\right]},\]
\[\text{and,}\quad {\epsilont_{ij}^{(p)} = {\Br{2\log\Paren{\frac{1.25\;p_{ij}}{\delta - p_{ij}\delta_j'}}}^{\frac{1}{2}}\frac{\alpha_{ij}{\rm R}}{\sqrt{\zetao_j + \sigma_{jj}^2 - r_j}}}}.\]

Moreover, for any $\uv, \uv' \in \mathbb{B}_d(\Rm)$,
\begin{align}
\label{eq:central_privacy_PS_local_data}
    &\Pr(\{\yv_j\}_{j \in [n]} \in \Scal^n \mid \xv_i = \uv) \nonumber\\
    & \qquad\leq e^{\epsilont_d}\Pr(\{\yv_j\}_{j \in [n]} \in \Scal^n \mid \xv_i = \uv') + \deltat_d,
\end{align}
where,
\[\epsilont_d = \sum_{\substack{j \in [n] \\ p_{ij} > 0}}\epsilont_{ij}^{(d)}, \quad \deltat_d = \delta\sum_{j \in [n]}p_j, \quad \delta \in \left(0, {\rm min}_{\substack{j \in [n] \\ p_{ij} > 0}}\{p_{ij}\}\right],\]
\[\text{and,}\quad \epsilont_{ij}^{(d)} = {\Br{2\log\Paren{\frac{1.25\;p_{ij}}{\delta - p_{ij}\delta_j'}}}^{\frac{1}{2}}\frac{2\alpha_{ij}{\rm R}}{\sqrt{\zetao_j + \sigma_{jj}^2 - r_j}}}.\]
\end{theorem}

Here, for any $\delta_j$, the parameter $r_j$ can be found using Corollary \ref{cor:choice_of_t_bernstein_inequality}.
Note that compared to $\epsilon_{ij}^{(p)}$ and $\epsilon_{ij}^{(d)}$, the expressions for $\epsilont_{ij}^{(p)}$ and $\epsilont_{ij}^{(d)}$ have an additional $\sigma_{jj}^2$ in the denominator.
This is a consequence of the fact that after aggregating $\xtvu_j$ from its neighbors, node $j$ adds privacy noise to its own data before transmitting to the PS.

\section{Privacy-constrained weight and variance optimization}
\label{sec:privacy-constrained-weight-optimization}

When deriving the utility-privacy tradeoff, our objective is to minimize the MSE at the PS subject to desired privacy guarantees, namely $(\underline{\epsilon}_{ij},\underline{\delta}_{ij}p_{ij})$ node-node local differential privacy.
Here, $\underline{\epsilon}_{ij}, \underline{\delta}_{ij}$ are pre-specified parameters that quantify the extent to which node $i$ trusts the link $i \to j$ against an eavesdropper.
While finding the optimal collaboration scheme, it is important  to consider the fact that due to relaying, a specific node $i$'s data is received via multiple nodes at the PS.
More precisely, the expected bias in the contribution of $\xv_i$ to the estimated mean ($\xvoh$) is given by $(S_i- 1)\xv_i$, and the cumulative bias from all the nodes is $\beta(\pv,\Pv,\Av) = \sum_{i \in [n]}\left\lvert S_i - 1 \right\rvert$.
Consequently, we aim to solve the following bias-regularized MSE minimization problem:
\begin{align}
    \label{eq:weight-optimization}
    &\min_{\Av,\Sigmav}\; \sigma_{\rm tiv}^2(\pv, \Pv, \Av) + \sigma_{\rm pr}^2(\pv,\Pv,\Sigmav) + \lambda\beta(\pv,\Pv, \Av) \nonumber\\
    &\text{ s.t. } \Br{2\log\Paren{\frac{1.25 }{\underline{\delta}_{ij}}}}^{\frac{1}{2}}\frac{2\alpha_{ij}{\rm R}}{\sigma_{ij}}\leq \underline{\epsilon}_{ij}, \; \alpha_{ij}\geq 0, \; \sigma_{ij} \geq 0.
\end{align}


Here, the constraints require that the collaboration weights and privacy noise be non-negative, and ensure that the peer-to-peer privacy constraints are respected.
Since the privacy constraints are cone constraints (denoted as $\Kcal$ in Fig. \ref{fig:weight-optimization-cone-constraints}), and are separable with respect to the optimization variables $(\alpha_{ij}, \sigma_{ij})$, we apply projected gradient descent to solve \eqref{eq:weight-optimization}.
%
%
Furthermore, $\lambda$ controls the strength of the regularization and we study its impact in \S \ref{sec:numerical-simulations}.
Instead of $\ell_1$, we can also choose an $\ell_2$-regularization term to control the bias, i.e., $\beta(\pv, \Pv, \Av) = \sum_{i \in [n]}\Paren{S_i - 1}^2$ (ref. to \S \ref{subsec:erdos-renyi-collaboration}).

\subsection{Projected gradient descent for MSE minimization}
\label{subsec:projected-gradient-descent}

Let us denote $\beta_{ij} \triangleq \frac{2\Rm}{\underline{\epsilon}_{ij}}\Br{2\log\Paren{\frac{1.25}{\underline{\delta}_{ij}}}}^{\frac{1}{2}}$.
For each $i,j \in [n]$, we start with a feasible iterate $\Paren{\alpha_{ij}^{(0)}, \sigma_{ij}^{(0)}}$ that satisfies $\alpha_{ij}^{(0)} \geq 0$, $\sigma_{ij}^{(0)} \geq \beta_{ij}\alpha_{ij}^{(0)}$, and take gradient steps followed by projections onto the cone, which can be computed in closed form.
The Pseudocode for the weight and noise variance optimization algorithm is provided in Alg. \ref{algo:projected-gradient-descent}, and the projection expressions are derived in App. \ref{app:weight-optimization-projected-gradient-descent}.
From \cite{boyd2004convex}, we know that the iterates of projected gradient descent for a Lipschitz objective converges to an $\varepsilon$-stationary point within $\mathrm{O}\Paren{\varepsilon^{-2}}$ iterations.\footnote{We choose projected gradient descent for its simplicity and ease of implementation. Other algorithms can also be used to solve \eqref{eq:weight-optimization}.}
In \S \ref{subsec:erdos-renyi-collaboration} below, we show that for a simple network topology, the solution of \eqref{eq:weight-optimization} can be obtained in closed form.

\begin{figure}[!htb]
\vspace{-5mm}
\centering
\begin{minipage}{0.8\linewidth}
\begin{algorithm}[H]
    \caption{{\bf \pricer}: Weight and noise variance optimization with projected gradient descent}
    \label{algo:projected-gradient-descent}
    \begin{algorithmic}[H]
    \State {\bf Input}: Feasible initial $\Av^{(0)}$ and $\Sigmav^{(0)}$.\\
    Maximal number of iterations: $T$
    \State {\bf Output}: Weight matrix $\Av_* = \Av^{(T)}$ and noise variances $\Sigmav_* = \Sigmav^{(T)}$ that minimizes \eqref{eq:weight-optimization}.
    \For{$t=1$ to $T$ for all $i,j \in [n]$}
    \State Take gradient descent steps:
    \begin{align*}
        &\widetilde{\alpha}_{ij} \gets \alpha_{ij}^{(t-1)} - \eta_{\alpha}\Paren{\frac{\partial \Ecal}{\partial \alpha_{ij}}}_{\Av^{(t-1)}} \nonumber\\
        \text{and, }\quad &\widetilde{\sigma}_{ij} \gets \sigma_{ij}^{(t-1)} - \eta_{\sigma}\Paren{\frac{\partial \Ecal}{\partial \sigma_{ij}}}_{\Sigmav^{(t-1)}}
    \end{align*}

    \State Project onto cone constraints:
    \begin{align*}
        &\Paren{\alpha_{ij}^{(t)}, \sigma_{ij}^{(t)}}  \\
        =
        &\begin{cases}
        \hspace{1cm}\Paren{0, \widetilde{\sigma}_{ij}} \quad \text{ if } \; \widetilde{\sigma}_{ij} \geq 0 \text{ and } \widetilde{\alpha}_{ij} < 0, \\
        \Paren{\Paren{\frac{\widetilde{\alpha}_{ij} + \beta_{ij}\widetilde{\sigma}_{ij}}{1 + \beta_{ij}^2}}_+, \Paren{\frac{\beta_{ij}(\widetilde{\alpha}_{ij} + \beta_{ij}\widetilde{\sigma}_{ij})}{1 + \beta_{ij}^2}}_+} \nonumber\\
        \qquad \text{ if } \widetilde{\sigma}_{ij} <0 \text{ or } \left\{ \widetilde{\sigma}_{ij} \geq 0 \text{ and } \widetilde{\sigma}_{ij} < \beta_{ij}\widetilde{\alpha}_{ij}\right\},\\
        \hspace{1cm}\Paren{\widetilde{\alpha}_{ij}, \widetilde{\sigma}_{ij}} \quad \text{ otherwise}.
        \end{cases}
    \end{align*}
    \EndFor
    \end{algorithmic}
\end{algorithm}               
\end{minipage}
\vspace{-5mm}
\end{figure}

\subsection{Optimal collaboration over an Erdös-Rényi Topology}
\label{subsec:erdos-renyi-collaboration}

\begin{figure*}[b]
\hrulefill 
\vspace*{4pt} 
\begin{align}
\label{eq:optimal_weights_erdos-renyi}
    \alpha_*(\lambda) &= q\left(\frac{\Rm^2}{\lambda n^2}\Paren{1-p + \frac{d\xi^2}{\epsilon^2}}\left(\frac{\Rm^2}{\lambda n^2}(1 + (m-1)q) + q\right) + mpq\Paren{\frac{\Rm^2}{\lambda mn}(1 + (m-1)q) + q}\right)^{-1}\Paren{\frac{\Rm^2}{\lambda n} + 1}, \nonumber \\
    \gamma_*(\lambda) &= \left(\frac{\Rm^2}{n^2}\left(1 + (m-1)q\right) + \lambda q\right)^{-1}\left(\frac{\Rm^2}{n^2}\Paren{n - (1 + (m-1)q)(n-m)p\alpha_*(\lambda)} + \lambda\right) \quad\text{and,} \quad \sigma_*(\lambda) = \xi\Rm\frac{\alpha_*(\lambda)}{\epsilon}.
\end{align}
\vspace*{4pt} 
\end{figure*}

Consider a setup of $n$ nodes that can collaborate with each other over an Erdös-Rényi graph, i.e., $p_{ij} = p$ for $i \neq j$ and $p_{ii} = 1$ for all $i,j \in [n]$.
Furthermore, suppose only a subset of these nodes, denoted by $\Mcal \subseteq [n]$ with cardinality $|\Mcal| = m$, can communicate with the PS reliably, i.e., $p_i = q$ if $i \in \Mcal$, and $p_i = 0$ otherwise.
Moreover, $(\epsilon_{ij}, \delta_{ij}) = (\epsilon, \delta)$ for all $i \in \Mcalo, j \in \Mcal$, where $\Mcalo = [n] \setminus \Mcal$.

From symmetry, for some $\alpha, \gamma > 0$, we have $\alpha_{ii} = \gamma$ for $i \in [n]$, $\alpha_{ij} = \alpha$ for $j \in \Mcal, i \in \Mcalo\setminus\{j\}$, and $\alpha_{ij} = 0$.
Furthermore, $\sigma_{ii} = 0$ as each node trusts itself, and $\sigma_{ij} = \sigma$ whenever $j \neq i$ and $\alpha_{ij} > 0$.
We consider $\beta(\pv, \Pv, \Av) = \sum_{i \in [n]}\Paren{S_i - 1}^2$ for analytical tractability.
We show in App. \ref{app:collaboration-over-erdos-renyi-topologies} that the solution of \eqref{eq:weight-optimization} is given by \eqref{eq:optimal_weights_erdos-renyi} below.
When $\lambda \to \infty$, we get $\alpha_* = (mpq)^{-1}$ and $\gamma_* = q^{-1}$.
With these weights,
\begin{align}
\label{eq:mse_erdos-renyi_graph_colab}
    {\rm MSE}_* = \Rm^2\left[\frac{(n-m)}{n^2pq}\Paren{\frac{1-p}{m} + \frac{\xi^2d}{\epsilon^2}} + \frac{1-q}{mq}\right].
\end{align}

In order to compare with a setup with no collaboration, consider $p_i = q$ for $i \in \Mcal$, and $p_i = q'$ for $i \in \Mcalo$, where $q' \to 0$.
In the absence of collaboration, in order to maintain an unbiased estimate the PS, node $i$ should transmit $p_i^{-1}\xv_i$ to the PS.
In this case, there are no peer-to-peer privacy concerns.
However, the MSE, which is given by
\begin{align*}
    \frac{1}{n^2}\Ebb\norm{\sum_{i \in [n]} \Paren{\tau_i\frac{\xv_i}{p_i} - \xv_i}}^2 = \frac{\Rm^2}{n^2}\Paren{\frac{m}{q} + \frac{n-m}{q'} - n},
\end{align*}
increases to $\infty$ as $q' \to 0$.
On the other hand, \eqref{eq:mse_erdos-renyi_graph_colab} shows that the MSE with collaboration is finite.

{\bf Central privacy guarantees for PriCER of Erdös-Rényi topology}: In App. \ref{app:collaboration-over-erdos-renyi-topologies}, we show that when $m < n$ and $p > \frac{7}{8}$, the privacy leakage at the relay for preserving anonymity of the identity of a participating node, as defined in Thm. \ref{thm:central_privacy_node_identity_protection} scales as $\epsilon_{ij}^{(p)} = \Om\Paren{\frac{1}{m\sqrt{n-m}}}$.
Moreover, from Thm. \ref{thm:central_privacy_node_dataset_protection}, since $\epsilon_{ij}^{(d)} = 2\epsilon_{ij}^{(p)}$, the same scaling also holds for privacy leakage for protecting the local data.

\section{Numerical simulations}
\label{sec:numerical-simulations}

\begin{figure}[h]
    \includegraphics[width=0.96\columnwidth, trim={0cm 3cm 0cm 3.5cm}, clip]{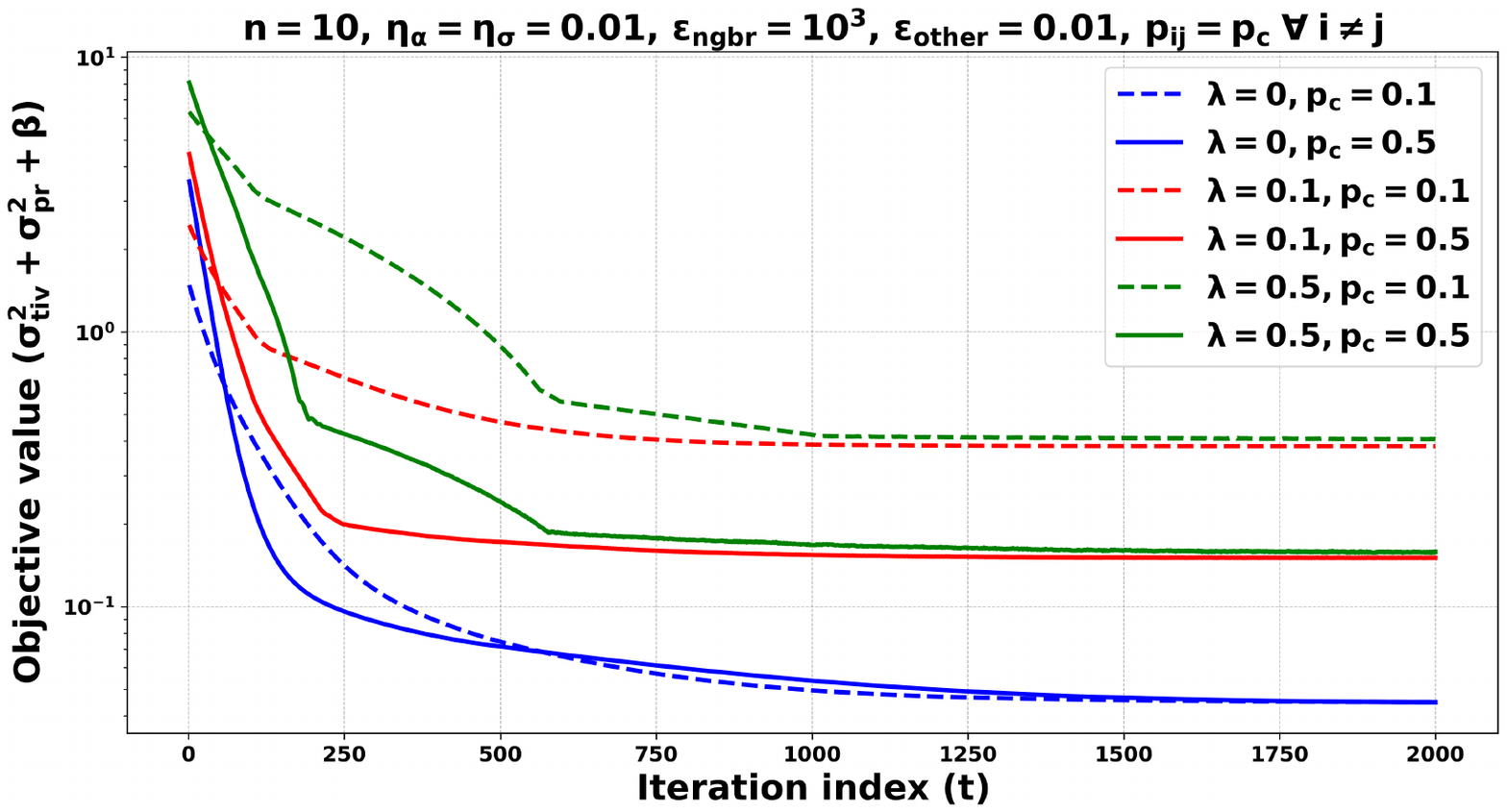}
    \captionsetup{font=small}
    \caption{Projected gradient descent with $2$ trustworthy neighbors}
    \label{fig:convergence_proj_gradient_descent}
\end{figure}

{\bf Convergence of Alg. \ref{algo:projected-gradient-descent}}: The objective function of \eqref{eq:weight-optimization} is, in general, non-convex.
In Fig. \ref{fig:convergence_proj_gradient_descent}, we see that projected gradient descent with constant learning rate $\eta_{\alpha} = \eta_{\sigma} = 0.01$, converges in the cases shown.
This value of learning rate has been chosen after tuning it over the set $[0.001, 0.01, 0.1]$.
To obtain these plots, $n = 10$ nodes are considered with connectivities to the PS given by $\pv = [0.1, 0.1, 0.8, 0.1, 0.1, 0.9, 0.1, 0.1, 0.9, 0.1]$.
Note that some of these probabilities are good, some mediocre, and some quite low. 
All plots are averaged over $4$ independent random initializations for gradient descent.
Furthermore, the nodes are connected to each other according to an Erd\H{o}s-R\'enyi topology, where for all $i,j \in [n]$, the edge between nodes $i$ and $j$ is present with probability $p_{ij} = p_c = 0.9$.
Nodes only trust their immediate $1$-hop neighbors, and privacy parameters are chosen accordingly as follows: For every $i \in [n]$, $\underline{\epsilon}_{ij} = \varepsilon_{ngbr} = 10^3$ if $j = i$, $j = (i + 1) \;{\rm mod}\; n$, or $j = (i-1) \;{\rm mod}\; n$, and $\underline{\epsilon}_{ij} = \varepsilon_{other} = 1$ otherwise.
$\delta_{ij} = 10^{-3}$ for all $i,j \in [n]$.
The ensures that there is effectively no privacy constraint for nodes communicating with their neighbors.

As expected, with higher values of the bias regularization parameter $\lambda$, the optimal MSE value at convergence increases. 
This leads to a {\it bias-MSE tradeoff} between the optimized MSE, i.e., $\sigma_{\rm tiv}^2(\pv, \Pv, \Av_*) + \sigma_{\rm pr}^2(\pv, \Pv, \Sigmav_*)$, and the total bias at the PS, i.e., $\beta(\pv, \Pv, \Av_*)$, where $\Av_*$ and $\Sigmav_*$ denote the optimized weights and privacy noise variance at convergence.
We study this tradeoff in Table \ref{tab:bias-mse-tradeoff}, which shows the total bias and the optimized MSE, respectively.
We sweep both the bias regularization parameter $\lambda$ and the node-node connectivity probability $(p_c)$ for the Erd\H{o}s-R\'enyi topology.

\begin{figure*}[t]
  \begin{minipage}[t]{0.32\textwidth}
    \centering
    \includegraphics[width=\textwidth, trim={0cm 2cm 0cm 2cm}, clip]{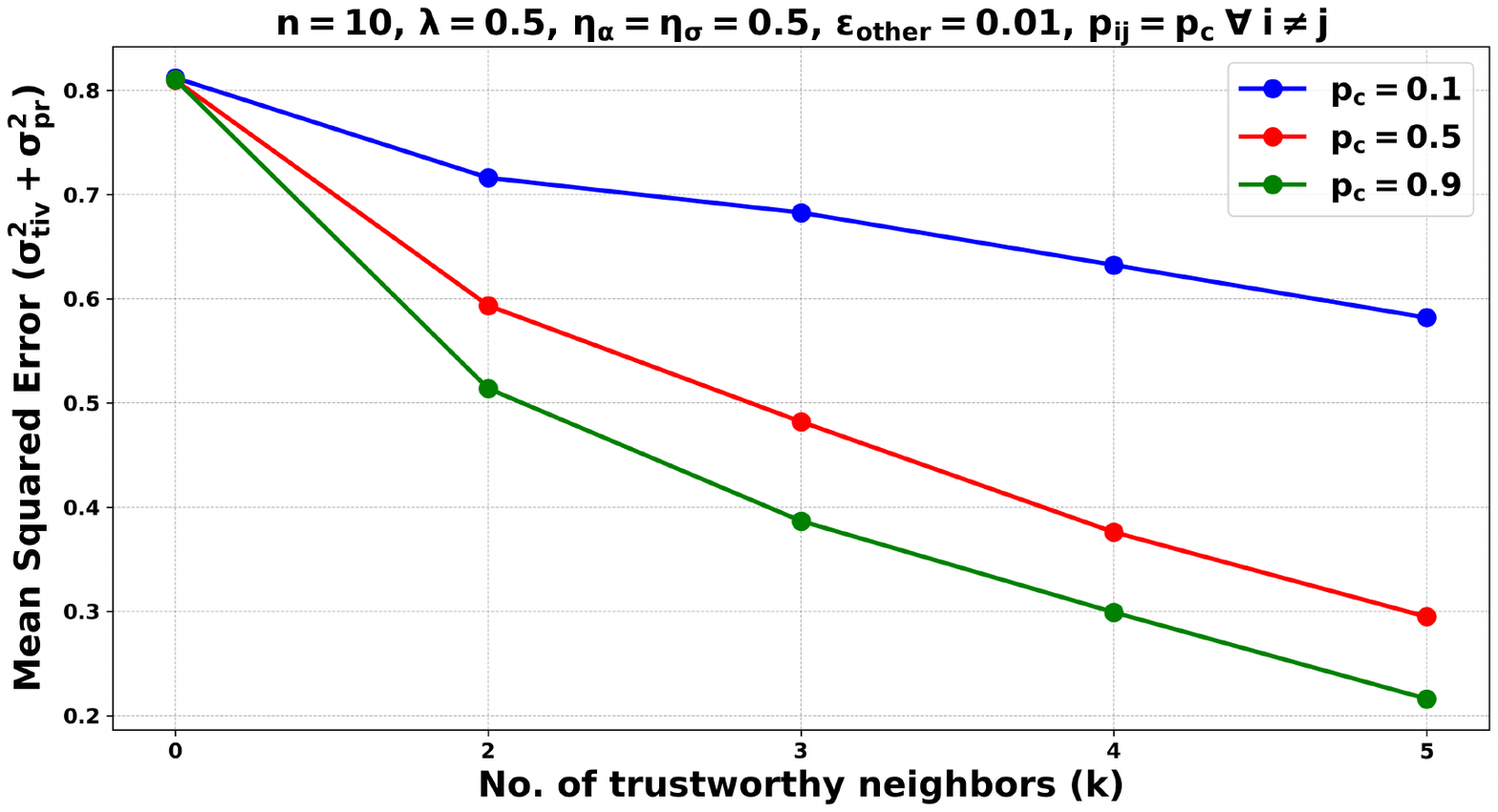}
    \captionsetup{justification=centering, font=small}
    \caption{Variation of MSE with number of trustworthy neighbors}
    \label{fig:mse_variation_trustworthy_ngbrs}
  \end{minipage}
  \hfill
  \begin{minipage}[t]{0.32\textwidth}
    \centering
    \includegraphics[width=\textwidth, trim={0cm 2cm 0cm 2cm}, clip]{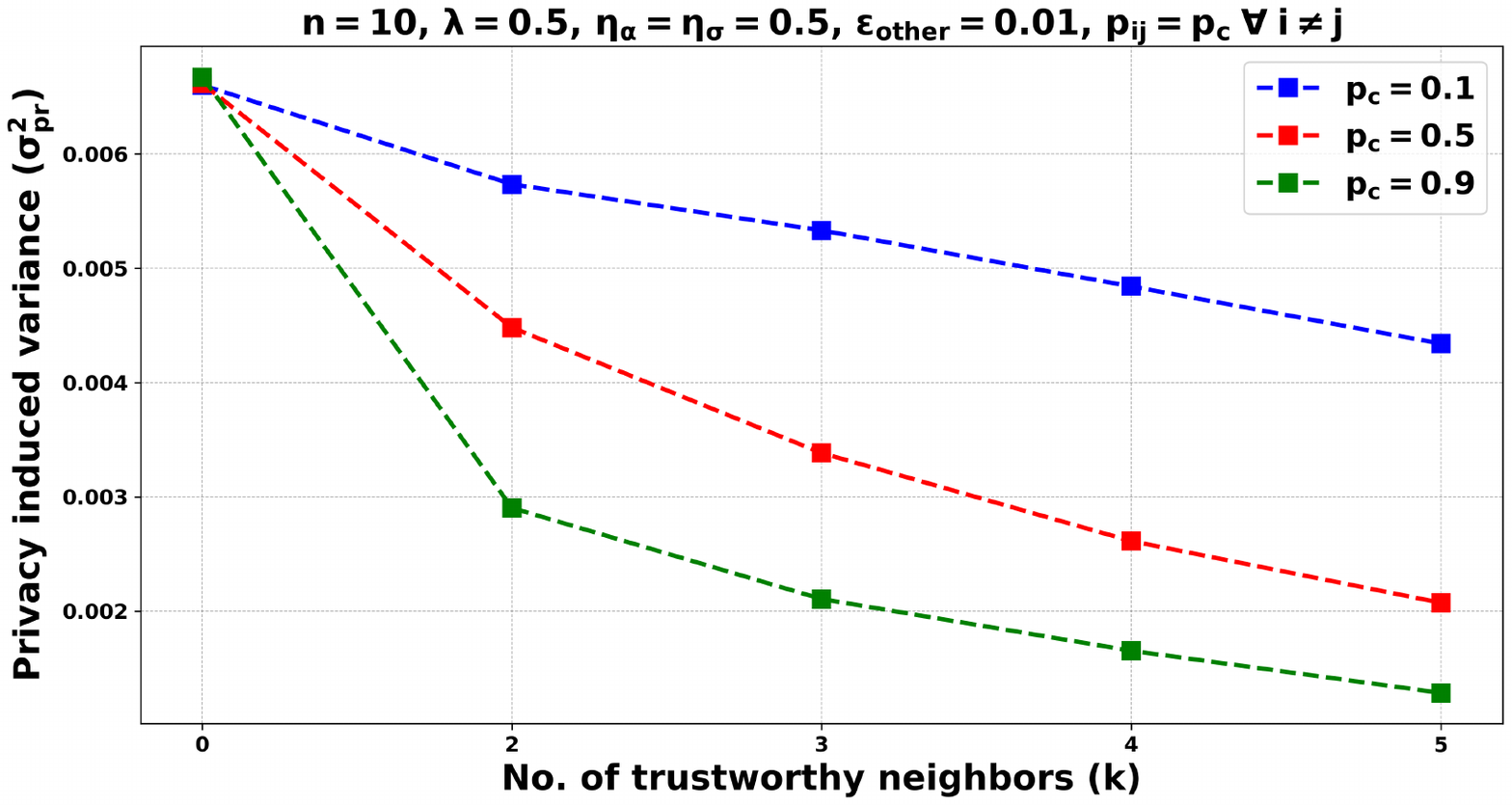}
    \captionsetup{justification=centering, font=small}
    \caption{Variation of PIV with number of trustworthy neighbors}
    \label{fig:piv_variation_trustworthy_ngbrs}
  \end{minipage}
  \hfill
  \begin{minipage}[t]{0.33\textwidth}
    \centering
    \includegraphics[width=0.75\textwidth, height=0.12\textheight, trim=0cm 0cm 0cm 1.1cm, clip]{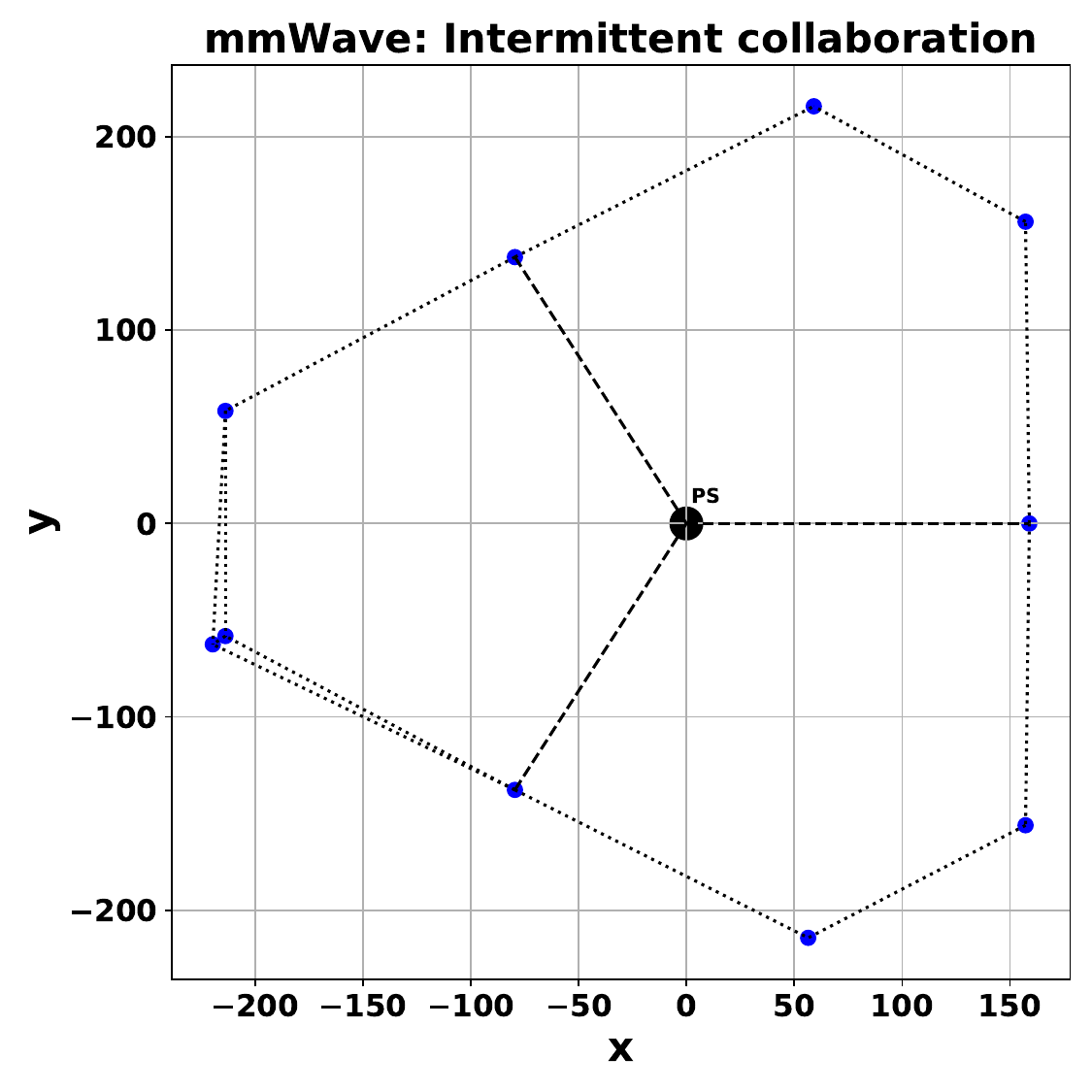}
    \captionsetup{justification=centering, font=small}
    \caption{Topology for $\rm K$-means clustering. $p_i$ and $p_{ij}$ are determined according to \cite{akdeniz_mmWave}.}
    \label{fig:scattered_topology}
  \end{minipage}
\end{figure*}

\begin{table}[h]
\captionsetup{justification=centering, font=small}
\captionof{table}{Bias - MSE tradeoff}
\label{tab:bias-mse-tradeoff}
    \centering
    \begin{tabular}{|c|c|c|c|}
        \hline
        \multicolumn{4}{|c|}{\scriptsize \textbf{Total bias} $\beta(\pv, \Pv, \Av_*)$} \\
        \hline
        {\centering \small $p_c $ $\downarrow$ $\lambda$ $\rightarrow$} & 
        {\centering \small $0$} & 
        {\centering \small $0.1$} & 
        {\centering \small $0.5$}\\
        \hline
        {\centering \small $0.1$} & 
        {\centering \small $12.799$} & 
        {\centering \small $0.4125$} & 
        {\centering \small $\bf 0.0025$} \\
        \hline
        {\centering \small $0.5$} & 
        {\centering \small $12.122$} & 
        {\centering \small $0.0082$} & 
        {\centering \small $\bf 0.0020$} \\
        \hline
        \multicolumn{4}{c}{} \\
        \hline
        \multicolumn{4}{|c|}{\scriptsize \textbf{MSE} $\sigma^2_{\rm tiv}(\pv, \Pv, \Av_*) + \sigma^2_{\rm pr}(\pv, \Pv, \Sigmav_*)$} \\
        \hline
        {\centering \small $p_c $ $\downarrow$ $\lambda$ $\rightarrow$} & 
        {\centering \small $0$} &
        {\centering \small $0.1$} & 
        {\centering \small $0.5$} \\
        \hline
        {\centering \small $0.1$} & 
        {\centering \small $\bf 0.0449$} & 
        {\centering \small $0.3422$} & 
        {\centering \small $0.4039$}\\
        \hline
        {\centering \small $0.5$} & 
        {\centering \small $\bf 0.0448$} & 
        {\centering \small $0.1493$} & 
        {\centering \small $0.1538$} \\
        \hline
    \end{tabular}
\end{table}

{\bf Variation with number of trustworthy neighbors}: Next, we study how increased collaboration can help in reducing the MSE. 
In Fig. \ref{fig:mse_variation_trustworthy_ngbrs}, we consider a topology in which $n = 10$ nodes are arranged as a ring (as before), and each node trusts $k$-hop neighbors, where we vary $k$ along the X-axis.
Furthermore, we consider that only one node has good connectivity to the PS, i.e., $p_i = 0.9$ if $i = 0$, otherwise $p_i = 0.1$.
This setup ensures that collaboration is critical to ensure a small MSE.
We set $\underline{\epsilon}_{ij} = \varepsilon_{other}$ whenever a node does not trust another node
As expected, the MSE decreases as collaboration amongst trustworthy neighbors (i.e., $k$) increases.
As opposed to the settings in prior works \cite{yemini2022_ISIT_robust, saha2022colrel, yemini2022robust_TWC}, here, the collaboration is not limited by a lack of communication opportunity between the nodes, but rather, their unwillingness to communicate with other (untrustworthy) nodes.
Consequently, even when node-node connectivity is good, i.e., $p_c = 0.9$, the optimized collaboration weight $\alpha_{ij}^*$ is small when node $i$ does not trust node $j$.
Moreover, while the above-mentioned works show that collaboration amongst nodes helps in reducing the topology induced variance due to poor connectivity within the network, we show that increased collaboration also helps in reducing the privacy induced variance. 
This is shown in Fig. \ref{fig:piv_variation_trustworthy_ngbrs}, where a higher value of $k$ implies that a node does not need to perturb its data too much while sharing it with its neighbors it trusts, resulting in a smaller PIV.

{\bf Distributed $\rm K$-means clustering}: We also evaluate our algorithm on the task of clustering a dataset distributed across $n$ nodes into $\rm K $ clusters. 
Each cluster is represented by its centroid and any datapoint is clustered to its nearest centroid.
In distributed $\rm K$-means clustering, each node first computes $\rm K$ local centroids. 
On each node, this is done by initializing the centroids according to the $\rm K$-means++ algorithm \cite{kmeans_plus_plus}, clustering the local datapoints to their nearest centroid, recomputing the cluster means, and repeating this process.
Once the local centroids have been computed on each node, they are transmitted to the central PS over intermittently failing communication links. 
The PS computes the average of these local centroids to obtain the global centroids, broadcasts them to each of the nodes.
Each node then uses these centroids as their new initialization and repeats the local clustering process.
This process is repeated over multiple iterations.
We choose relative inertia as the evaluation metric for the clustering performance.
Inertia is defined as the sum of squared distances of each datapoint $\xv_i$ from its closest cluster $c(x_i)$.
Relative inertia is defined as the ratio of the inertia of our distributed $\rm K$-means algorithm and the inertia of the centralized variant, wherein the entire dataset is assumed to be present on a single node.
Mathematically, $\text{Inertia (relative)} \triangleq \text{Inertia (distributed)}/\text{Inertia (centralized)}$ where,
\begin{equation}
    \label{def:relative_inertia}
    \text{Inertia} \triangleq \sum_{i \;\in\; \text{dataset}}\left(x_i - c(\xv_i)\right)^2
\end{equation}

We consider clustering two datasets: {\rm (i)} Raw CIFAR-10 images \cite{cifar10}, and $\rm (ii)$ Fashion-MNIST \cite{xiao2017fashionmnist} image embeddings extracted using pre-trained MobileNet-V3 \cite{howard2017mobilenets}.
We consider $n = 10$ nodes, $\rm K = 10$ clusters, $10$ communication rounds with the PS and $5$ local iterations in each round, over three setups: $\rm (i)$ sole good node (i.e., as before, only one node has $p_i = 0.9$ and all other node have $p_i = 0.1$) and no collaboration, $\rm (ii)$ sole good node with $6$ trustworthy neighbors which enables collaboration via PriCER with parameters $(p_{ij} = 0.9, \epsilon_{ij} = 0.01, \delta = 10^{-3})$, $\rm (iii)$ a scattered topology as shown in Fig. \ref{fig:scattered_topology}, where nodes do not collaborate with each other, and $\rm (iv)$ the same scattered topology with PriCER collaboration with parameters $(\epsilon_{ij} = 0.01, \delta = 10^{-3})$.
The connectivity probabilities between nodes or between the node and the PS is a function of distance  ($\rm \delta$), given by $p = \min\left(1, \mathrm{e}^{-\delta/30 + 5.2}\right)$.
This expression is used to model the outage probability of mmWave communication links \cite[Eq. 8a]{akdeniz_mmWave}, and our scattered topology is visualized in Fig. \ref{fig:scattered_topology}.
We assume that two nodes $i$ and $j$ trust each other if $p_{ij} > 0.5$, i.e., they have a relatively high probability of connectivity.
From Tab. \ref{tab:kmeans_clustering_expts}, we can see that the relative inertia for clustering is smaller when the nodes collaborate with each other.

\begin{table}[]
\captionsetup{font=small}
\caption{Relative inertia of $\rm K$-means clustering for different datasets over ``sole good node" and ``scattered"  network topologies. Smaller values indicate better performance}
\label{tab:kmeans_clustering_expts}
    \centering
    \begin{tabular}{|c|p{2.0cm}|p{2.1cm}|}
    \hline
    {\bf Setup} $\downarrow$ {\bf Dataset} $\rightarrow$ & \makecell{CIFAR-10 \\ (raw images)} & \makecell{FMNIST \\ (MobileNet-V3)} \\
    \hline
    Sole good node (no colab) & $2.101 \pm 0.007$ & $6.226 \pm 3.054$ \\
    \hline
    {\bf Sole good node ({\sc \textbf{PriCER}})} & $\mathbf{1.689 \pm 0.030}$ & $\mathbf{4.168 \pm 2.474}$ \\
    \hline
    Scattered (no colab) & $4.454 \pm 0.758$ & $6.593 \pm 0.626$ \\
    \hline
    {\bf Scattered ({\sc \textbf{PriCER}})} & $\mathbf{1.262 \pm 0.019}$ & $\mathbf{1.315 \pm 0.085}$ \\
    \hline
  \end{tabular}
\end{table}

\section{Conclusions}
\label{sec:conclusions}

We have considered the problem of mean estimation over intermittently connected networks with collaborative relaying subject to differential privacy constraints.
The nodes participating in the collaboration do not trust each other completely and, in order to ensure privacy, they scale and perturb their local data when sharing with others.
We have shown that local and central privacy can be guaranteed by leveraging the different sources of stochasticity in our setup, namely, the Gaussian perturbations added by a node before collaborating, as well as the intermittent nature of the node-node and node-PS connectivities. 
As the MSE of the mean estimated at the PS was comprised of the topology induced variance and the privacy induced variance, we have proposed a two-stage consensus algorithm ({\sc PriCER}) that jointly optimizes the scaling weights and perturbation noise variance so as to minimize the MSE while respecting the peer-to-peer privacy constraints.
This joint optimization also introduces a tradeoff between the MSE and the bias of the estimated mean, and we traverse this tradeoff curve by tuning the bias regularization hyperparameter.
We have performed numerical simulations and shown how {\sc PriCER} can do collaborative, yet private, mean estimation over intermittently connected networks, and that it outperforms non-collaborative strategies.

\let\oldbibliography\thebibliography
\renewcommand{\thebibliography}[1]{\oldbibliography{#1}
\setlength{\itemsep}{0pt}} 

\bibliographystyle{IEEEtran}
\bibliography{refs}

\begin{thebibliography}{10}
\providecommand{\url}[1]{#1}
\csname url@samestyle\endcsname
\providecommand{\newblock}{\relax}
\providecommand{\bibinfo}[2]{#2}
\providecommand{\BIBentrySTDinterwordspacing}{\spaceskip=0pt\relax}
\providecommand{\BIBentryALTinterwordstretchfactor}{4}
\providecommand{\BIBentryALTinterwordspacing}{\spaceskip=\fontdimen2\font plus
\BIBentryALTinterwordstretchfactor\fontdimen3\font minus \fontdimen4\font\relax}
\providecommand{\BIBforeignlanguage}[2]{{%
\expandafter\ifx\csname l@#1\endcsname\relax
\typeout{** WARNING: IEEEtran.bst: No hyphenation pattern has been}%
\typeout{** loaded for the language `#1'. Using the pattern for}%
\typeout{** the default language instead.}%
\else
\language=\csname l@#1\endcsname
\fi
#2}}
\providecommand{\BIBdecl}{\relax}
\BIBdecl

\bibitem{saha2023isit-p2p-privacy}
R.~Saha, M.~Seif, M.~Yemini, A.~J. Goldsmith, and H.~V.~Poor, ``Collaborative mean estimation over intermittently connected networks with peer-to-peer privacy,'' in \emph{Proceedings of the 2023 IEEE International Symposium on Information Theory}, 2023, pp. 174--179.

\bibitem{mcmahan2017communication}
B.~McMahan, E.~Moore, D.~Ramage, S.~Hampson, and B.~A. y~Arcas, ``Communication-efficient learning of deep networks from decentralized data,'' in \emph{Proceedings of the 20th International Conference on Artificial Intelligence and Statistics}.\hskip 1em plus 0.5em minus 0.4em\relax PMLR, 2017, pp. 1273--1282.

\bibitem{balcan2013distributed}
M.-F.~F. Balcan, S.~Ehrlich, and Y.~Liang, ``Distributed $ k $-means and $ k $-median clustering on general topologies,'' \emph{Advances in Neural Information Processing Systems}, vol.~26, p. 1995–2003, 2013.

\bibitem{suresh2017distributed}
A.~T. Suresh, X.~Y. Felix, S.~Kumar, and H.~B. McMahan, ``Distributed mean estimation with limited communication,'' in \emph{Proceedings of the International Conference on Machine Learning}.\hskip 1em plus 0.5em minus 0.4em\relax PMLR, 2017, pp. 3329--3337.

\bibitem{kar2008distributed}
S.~Kar and J.~M.~F. Moura, ``Distributed consensus algorithms in sensor networks with imperfect communication: Link failures and channel noise,'' \emph{IEEE Transactions on Signal Processing}, vol.~57, no.~1, pp. 355--369, 2009.

\bibitem{tandon2017gradient}
R.~Tandon, Q.~Lei, A.~G. Dimakis, and N.~Karampatziakis, ``Gradient coding: Avoiding stragglers in distributed learning,'' in \emph{Proceedings of the International Conference on Machine Learning}.\hskip 1em plus 0.5em minus 0.4em\relax PMLR, 2017, pp. 3368--3376.

\bibitem{kar_2008_topology_design}
S.~Kar and J.~M.~F. Moura, ``Sensor networks with random links: Topology design for distributed consensus,'' \emph{IEEE Transactions on Signal Processing}, vol.~56, no.~7, pp. 3315--3326, 2008.

\bibitem{kar_2012_nonlinear_observation_model}
S.~Kar, J.~M.~F. Moura, and K.~Ramanan, ``Distributed parameter estimation in sensor networks: Nonlinear observation models and imperfect communication,'' \emph{IEEE Transactions on Information Theory}, vol.~58, no.~6, pp. 3575--3605, 2012.

\bibitem{jhunjhunwala2021leveraging}
D.~Jhunjhunwala, A.~Mallick, A.~Gadhikar, S.~Kadhe, and G.~Joshi, ``Leveraging spatial and temporal correlations in sparsified mean estimation,'' \emph{Advances in Neural Information Processing Systems}, vol.~34, pp. 14\,280--14\,292, 2021.

\bibitem{gapeyenko_mobile_blockers}
M.~{Gapeyenko}, A.~{Samuylov}, M.~{Gerasimenko}, D.~{Moltchanov}, S.~{Singh}, M.~R. {Akdeniz}, E.~{Aryafar}, N.~{Himayat}, S.~{Andreev}, and Y.~{Koucheryavy}, ``On the temporal effects of mobile blockers in urban millimeter-wave cellular scenarios,'' \emph{IEEE Transactions on Vehicular Technology}, vol.~66, no.~11, pp. 10\,124--10\,138, Nov 2017.

\bibitem{yasamin}
Y.~{Yan} and Y.~{Mostofi}, ``Co-optimization of communication and motion planning of a robotic operation under resource constraints and in fading environments,'' \emph{{IEEE} Transactions on Wireless Communications}, vol.~12, no.~4, pp. 1562--1572, April 2013.

\bibitem{pappas}
M.~M. {Zavlanos}, M.~B. {Egerstedt}, and G.~J. {Pappas}, ``Graph-theoretic connectivity control of mobile robot networks,'' \emph{Proceedings of the {IEEE}}, vol.~99, no.~9, pp. 1525--1540, Sep. 2011.

\bibitem{zavlanos}
N.~Michael, M.~M. Zavlanos, V.~Kumar, and G.~J. Pappas, ``Maintaining connectivity in mobile robot networks,'' in \emph{Experimental Robotics}, 2009, pp. 117--126.

\bibitem{yemini_et_al:globecom2020}
M.~{Yemini}, S.~{Gil}, and A.~J. {Goldsmith}, ``Exploiting local and cloud sensor fusion in intermittently connected sensor networks,'' in \emph{Proceedings of the 2020 IEEE Global Communications Conference}, December 2020.

\bibitem{yemini_cloud_cluster}
M.~Yemini, S.~Gil, and A.~J. Goldsmith, ``Cloud-cluster architecture for detection in intermittently connected sensor networks,'' \emph{IEEE Transactions on Wireless Communications}, vol.~22, no.~2, pp. 903--919, 2023.

\bibitem{gu2021fast}
X.~Gu, K.~Huang, J.~Zhang, and L.~Huang, ``Fast federated learning in the presence of arbitrary device unavailability,'' in \emph{Advances in Neural Information Processing Systems}, 2021.

\bibitem{Yan2020DistributedNO}
Y.~Yan, C.~Niu, Y.~Ding, Z.~Zheng, F.~Wu, G.~Chen, S.~Tang, and Z.~Wu, ``Distributed non-convex optimization with sublinear speedup under intermittent client availability,'' \emph{ArXiv}, vol. abs/2002.07399, 2020.

\bibitem{wang_2022}
S.~Wang and M.~Ji, ``A unified analysis of federated learning with arbitrary client participation,'' in \emph{Advances in Neural Information Processing Systems}, vol.~35, 2022, pp. 19\,124--19\,137.

\bibitem{pmlr-v130-ruan21a}
Y.~Ruan, X.~Zhang, S.-C. Liang, and C.~Joe-Wong, ``Towards flexible device participation in federated learning,'' in \emph{Proceedings of the 24th International Conference on Artificial Intelligence and Statistics}, vol. 130, 2021, pp. 3403--3411.

\bibitem{anarchic_FL}
H.~Yang, X.~Zhang, P.~Khanduri, and J.~Liu, ``Anarchic federated learning,'' in \emph{Proceedings of the 39th International Conference on Machine Learning}, vol. 162, 2022, pp. 25\,331--25\,363.

\bibitem{Sun_2024}
Y.~Sun, Y.~Mao, and J.~Zhang, ``Mimic: Combating client dropouts in federated learning by mimicking central updates,'' \emph{IEEE Transactions on Mobile Computing}, p. 1–13, 2024.

\bibitem{wang2024a}
S.~Wang and M.~Ji, ``A lightweight method for tackling unknown participation statistics in federated averaging,'' in \emph{Proceedings of the 12th International Conference on Learning Representations}, 2024.

\bibitem{saha2022colrel}
\BIBentryALTinterwordspacing
R.~Saha, M.~Yemini, E.~Ozfatura, D.~Gunduz, and A.~Goldsmith, ``Colrel: Collaborative relaying for federated learning over intermittently connected networks,'' in \emph{Workshop on Federated Learning: Recent Advances and New Challenges (in Conjunction with NeurIPS 2022)}, 2022. [Online]. Available: \url{https://openreview.net/forum?id=8b0RHdh2Xd0}
\BIBentrySTDinterwordspacing

\bibitem{yemini2022_ISIT_robust}
M.~Yemini, R.~Saha, E.~Ozfatura, D.~G\"{u}nd\"{u}z, and A.~J. Goldsmith, ``Semi-decentralized federated learning with collaborative relaying,'' in \emph{Proceedings of the 2022 IEEE International Symposium on Information Theory}, 2022, pp. 1471--1476.

\bibitem{yemini2022robust_TWC}
M.~Yemini, R.~Saha, E.~Ozfatura, D.~Gündüz, and A.~J. Goldsmith, ``Robust semi-decentralized federated learning via collaborative relaying,'' \emph{IEEE Transactions on Wireless Communications (to appear)}, 2023.

\bibitem{dwork2014algorithmic}
C.~Dwork, A.~Roth \emph{et~al.}, ``The algorithmic foundations of differential privacy,'' \emph{Foundations and Trends{\textregistered} in Theoretical Computer Science}, vol.~9, no. 3--4, pp. 211--407, 2014.

\bibitem{kairouz2021advances}
P.~Kairouz, H.~B. McMahan, B.~Avent, A.~Bellet, M.~Bennis, A.~N. Bhagoji, K.~Bonawitz, Z.~Charles, G.~Cormode, R.~Cummings \emph{et~al.}, ``Advances and open problems in federated learning,'' \emph{Foundations and Trends{\textregistered} in Machine Learning}, vol.~14, no. 1--2, pp. 1--210, 2021.

\bibitem{balle2018privacy}
B.~Balle, G.~Barthe, and M.~Gaboardi, ``Privacy amplification by subsampling: Tight analyses via couplings and divergences,'' \emph{Advances in Neural Information Processing Systems}, vol.~31, 2018.

\bibitem{zhu2019poission}
Y.~Zhu and Y.-X. Wang, ``Poisson subsampled {R}{\'e}nyi differential privacy,'' in \emph{Proceedings of the International Conference on Machine Learning (ICML)}.\hskip 1em plus 0.5em minus 0.4em\relax PMLR, 2019, pp. 7634--7642.

\bibitem{luo2022tackling}
B.~Luo, W.~Xiao, S.~Wang, J.~Huang, and L.~Tassiulas, ``Tackling system and statistical heterogeneity for federated learning with adaptive client sampling,'' in \emph{Proceedings of the 2022 IEEE International Conference on Computer Communications}, 2022, pp. 1739--1748.

\bibitem{rizk2022federated}
E.~Rizk, S.~Vlaski, and A.~H. Sayed, ``Federated learning under importance sampling,'' \emph{IEEE Transactions on Signal Processing}, vol.~70, pp. 5381--5396, 2022.

\bibitem{balle2020privacy}
B.~Balle, P.~Kairouz, B.~McMahan, O.~Thakkar, and A.~Guha~Thakurta, ``Privacy amplification via random check-ins,'' \emph{Advances in Neural Information Processing Systems}, vol.~33, pp. 4623--4634, 2020.

\bibitem{asi_2022b}
H.~Asi, V.~Feldman, and K.~Talwar, ``Optimal algorithms for mean estimation under local differential privacy,'' in \emph{Proceedings of the 39th International Conference on Machine Learning}, vol. 162, 2022, pp. 1046--1056.

\bibitem{cummings_2022}
R.~Cummings, V.~Feldman, A.~McMillan, and K.~Talwar, ``Mean estimation with user-level privacy under data heterogeneity,'' in \emph{Advances in Neural Information Processing Systems}, vol.~35, 2022, pp. 29\,139--29\,151.

\bibitem{isik2023exact}
B.~Isik, W.-N. Chen, A.~Ozgur, T.~Weissman, and A.~No, ``Exact optimality of communication-privacy-utility tradeoffs in distributed mean estimation,'' in \emph{Proceedings of the 37th Conference on Neural Information Processing Systems}, 2023.

\bibitem{gaboardi19a}
M.~Gaboardi, R.~Rogers, and O.~Sheffet, ``Locally private mean estimation: $z$-test and tight confidence intervals,'' in \emph{Proceedings of the 22nd International Conference on Artificial Intelligence and Statistics}, vol.~89, 2019, pp. 2545--2554.

\bibitem{Xue2021MeanEO}
Q.~Xue, Y.~Zhu, and J.~Wang, ``Mean estimation over numeric data with personalized local differential privacy,'' \emph{Frontiers of Computer Science}, vol.~16, 2021.

\bibitem{oac1}
M.~M. {Amiri} and D.~{Gündüz}, ``Federated learning over wireless fading channels,'' \emph{IEEE Trans. Wireless Comms.}, vol.~19, no.~5, pp. 3546--3557, 2020.

\bibitem{oac2}
E.~Ozfatura, S.~Rini, and D.~Gündüz, ``Decentralized {SGD} with over-the-air computation,'' in \emph{IEEE Glob. Commun. Conf}, 2020.

\bibitem{oac3}
G.~Zhu, Y.~Wang, and K.~Huang, ``Broadband analog aggregation for low-latency federated edge learning (extended version),'' arxiv:1812.11494, 2019.

\bibitem{oac_privacy1}
B.~Hasircioglu and D.~Gunduz, ``Private wireless federated learning with anonymous over-the-air computation,'' arxiv:2011.08579, 2021.

\bibitem{oac_privacy2}
\BIBentryALTinterwordspacing
M.~Seif, W.-T. Chang, and R.~Tandon, ``Privacy amplification for federated learning via user sampling and wireless aggregation,'' arxiv:2103.01953, 2021. [Online]. Available: \url{https://ieeexplore.ieee.org/document/9518031}
\BIBentrySTDinterwordspacing

\bibitem{kairouz2015composition}
P.~Kairouz, S.~Oh, and P.~Viswanath, ``The composition theorem for differential privacy,'' in \emph{International conference on machine learning}.\hskip 1em plus 0.5em minus 0.4em\relax PMLR, 2015, pp. 1376--1385.

\bibitem{boyd2004convex}
S.~Boyd and L.~Vandenberghe, \emph{Convex Optimization}.\hskip 1em plus 0.5em minus 0.4em\relax Cambridge {U}niversity press, 2004.

\bibitem{akdeniz_mmWave}
M.~R. Akdeniz, Y.~Liu, M.~K. Samimi, S.~Sun, S.~Rangan, T.~S. Rappaport, and E.~Erkip, ``Millimeter wave channel modeling and cellular capacity evaluation,'' \emph{IEEE Journal on Selected Areas in Communications}, vol.~32, no.~6, pp. 1164--1179, 2014.

\bibitem{kmeans_plus_plus}
D.~Arthur and S.~Vassilvitskii, ``{K-Means++}: {T}he {A}dvantages of {C}areful {S}eeding,'' vol.~8, 01 2007, pp. 1027--1035.

\bibitem{cifar10}
\BIBentryALTinterwordspacing
A.~Krizhevsky, V.~Nair, and G.~Hinton, ``{CIFAR}-10 ({C}anadian {I}nstitute {F}or {A}dvanced {R}esearch).'' [Online]. Available: \url{http://www.cs.toronto.edu/~kriz/cifar.html}
\BIBentrySTDinterwordspacing

\bibitem{xiao2017fashionmnist}
H.~Xiao, K.~Rasul, and R.~Vollgraf, ``Fashion-{MNIST}: a novel image dataset for benchmarking machine learning algorithms,'' 2017.

\bibitem{howard2017mobilenets}
A.~G. Howard, M.~Zhu, B.~Chen, D.~Kalenichenko, W.~Wang, T.~Weyand, M.~Andreetto, and H.~Adam, ``Mobile{N}ets: Efficient convolutional neural networks for mobile vision applications,'' 2017.

\bibitem{mitzenmacher2017probability}
M.~Mitzenmacher and E.~Upfal, \emph{{P}robability and {C}omputing: {R}andomization and {P}robabilistic {T}echniques in {A}lgorithms and {D}ata {A}nalysis}.\hskip 1em plus 0.5em minus 0.4em\relax Cambridge {U}niversity press, 2017.

\end{thebibliography}

\appendices  

\vspace{-5mm}
\section{Estimation error analysis of \pricer: \\ Proof of Thm. \ref{thm:pricer-estimation-error-analysis}}
\label{app:pricer-estimation-error-analysis}

Note that the global estimate at the PS is
{\small
\begin{equation}
\xvoh = \frac{1}{n}\sum_{i \in [n]}\tau_i\sum_{j \in [n]}\tau_{ji}(\alpha_{ji}\xv_j + \nv_{ji}).
\end{equation}
}
Consequently, the MSE can be written as
{\small
\begin{align}
\label{eq:mse_decomposition_tiv_and_piv}
&\mathbb{E}\left\lVert\frac{1}{n}\sum_{i \in [n]}\tau_i\sum_{j \in [n]}\tau_{ji}(\alpha_{ji}\xv_j + \nv_{ji}) - \frac{1}{n}\sum_{i \in [n]}\xv_i\right\rVert_2^2 \nonumber\\
&= \underbrace{\mathbb{E}\left\lVert \frac{1}{n}\sum_{i \in [n]}\tau_i\sum_{j \in [n]}\tau_{ji}\alpha_{ji}\xv_j - \frac{1}{n}\sum_{i \in [n]}\xv_i  \right\rVert_2^2}_{\text{\rm Topology Induced Variance (TIV)}} \nonumber\\
&\qquad\qquad\qquad\qquad\qquad +\underbrace{\mathbb{E}\left\lVert \frac{1}{n}\sum_{i \in [n]}\tau_i\sum_{j \in [n]}\tau_{ji}\nv_{ji} \right\rVert_2^2}_{\text{\rm Privacy Induced Variance (PIV)}},
\end{align}
}

where the expectation is taken over the random connectivity and the local perturbation mechanism, and the equality follows because for any $j \in [n]$, the cross term is $\mathbb{E}\left[\sum_{i \in [n]}\tau_j\tau_{ij}\alpha_{ij}\xv_j^{\top}\mathbb{E}[\nv_{ij}]\right] = 0$.

The first term {\rm TIV} in \eqref{eq:mse_decomposition_tiv_and_piv} is solely affected by the intermittent connectivity of nodes.
The technique to upper bound it is similar to what is done in \cite[Thm. 3.2]{saha2022colrel}, except for the fact that we do not consider the unbiasedness constraint present in \cite{saha2022colrel}.
We allow for biased estimates at the PS and additional bias terms appear in the upper bound for TIV.
Nevertheless, the proof is presented here for the sake of completeness.
Recall our notation $S_i = \sum_{j \in [n]}p_jp_{ij}\alpha_{ij}$.
The TIV is then

{\small
\begin{align}
\label{eq:MSE_expression}
    &\mathbb{E}\norm{\frac{1}{n}\sum_{i \in [n]}\tau_i\sum_{j \in [n]}\alpha_{ji}\tau_{ji} \xv_j - \frac{1}{n}\sum_{i \in [n]}\xv_i}^2 \nonumber\\
    = &\frac{1}{n^2}\sum_{i \in [n]}\mathbb{E}\Br{\Paren{\sum_{j \in [n]}\tau_j\tau_{ij}\alpha_{ij} - 1}^2}\norm{\xv_i}^2 \nonumber\\
    + &\frac{1}{n^2}\sum_{\stackrel{i, l \in [n]:}{i \neq l}}\mathbb{E}\left[\Paren{\sum_{j \in [n]}\tau_j\tau_{ij}\alpha_{ij} - 1}\right.\nonumber\\
    &\qquad\qquad\qquad\left.\Paren{\sum_{m \in [n]}\tau_m\tau_{lm}\alpha_{lm}-1}\right]\xv_i^{\top}\xv_l.
\end{align}
}

In the first term of \eqref{eq:MSE_expression}, the coefficient of $\lVert \xv_i \rVert^2$ inside the $\sum_{i \in [n]}$ can be simplified as
{\small
\begin{align}
\label{eq:tiv_expression_first_term}    &\mathbb{E}\Br{\Paren{\sum_{j \in [n]}\tau_j\tau_{ij}\alpha_{ij} - 1}^2} \nonumber\\
    = &\sum_{j \in [n]} \mathbb{E}\Br{\tau_j^2\tau_{ij}^2\alpha_{ij}^2} + \sum_{\substack{j_1, j_2 \in[n]: \\ j_1 \neq j_2}} \mathbb{E}\Br{ \tau_{j_1}\tau_{j_2}\tau_{ij_1}\tau_{ij_2}\alpha_{ij_1}\alpha_{ij_2}} \nonumber\\
    &\qquad\qquad\qquad\qquad - 2\sum_{j \in [n]}\mathbb{E}\Br{\tau_j\tau_{ij}\alpha_{ij}} + 1 \nonumber\\
    &= \sum_{j \in [n]}p_jp_{ij}\alpha_{ij}^2 + \sum_{\substack{j_1, j_2 \in[n]: \\ j_1 \neq j_2}}p_{j_1}p_{j_2}p_{ij_1}p_{ij_2}\alpha_{ij_1}\alpha_{ij_2} \nonumber\\
    &\qquad\qquad\qquad\qquad - 2\sum_{j \in [n]}p_jp_{ij}\alpha_{ij} + 1 \nonumber\\
    &= \sum_{j \in [n]}p_jp_{ij}\alpha_{ij}^2 + \Paren{\sum_{j \in [n]}p_jp_{ij}\alpha_{ij}}^2 - \sum_{j \in [n]}p_j^2p_{ij}^2\alpha_{ij}^2 \nonumber\\
    &\qquad\qquad\qquad\qquad - 2\sum_{j \in [n]}p_jp_{ij}\alpha_{ij} + 1 \nonumber \\
    &= \sum_{j \in [n]}p_jp_{ij}\Paren{1 - p_jp_{ij}}\alpha_{ij}^2 + S_i(S_i - 2) + 1 \nonumber\\
    &= \sum_{j \in [n]}p_jp_{ij}\Paren{1 - p_jp_{ij}}\alpha_{ij}^2 + (S_i - 1)^2.
\end{align}
}

In the second term of \eqref{eq:MSE_expression}, the coefficient of $\xv_i^{\top}\xv_l$ inside the $\sum_{\substack{i,l \in [n] \\ i \neq l}}$ can be simplified as

{\small
\begin{align}
    &\sum_{j \in [n]}p_j p_{ij}p_{lj} \alpha_{ij} \alpha_{lj} + p_ip_l \Em_{\{i,l\}}\alpha_{il}\alpha_{li} + \sum_{\substack{m\in[n]: \\ m \neq l,i}}p_l p_m p_{il}p_{lm}\alpha_{il}\alpha_{lm} \nonumber \\
    &\hspace{2cm}+ \sum_{\substack{j\in[n]: \\ j \neq l}}\sum_{\substack{m\in[n]: \\ m \neq j}}p_j p_m p_{ij}p_{lm} \alpha_{ij}\alpha_{lm}-S_i -S_l + 1 \nonumber \\
    &= \sum_{j \in [n]}p_jp_{ij}p_{lj}\alpha_{ij}\alpha_{lj} + p_ip_l\Em_{\{i,l\}}\alpha_{il}\alpha_{li} \nonumber\\
    &\hspace{0.8cm}-S_i -S_l +1 + p_lp_{il}\alpha_{il}(S_l - p_lp_{ll}\alpha_{ll} - p_ip_{li}\alpha_{li}) \nonumber \\
     & \hspace{0.8cm}+ S_l(S_i - p_lp_{il}\alpha_{il}) - \sum_{j \in [n]}p_j^2p_{ij}p_{lj}\alpha_{ij}\alpha_{lj} + p_l^2p_{il}p_{ll}\alpha_{il}\alpha_{ll} \nonumber\\
     &= \sum_{j \in [n]}p_jp_{ij}p_{lj}\alpha_{ij}\alpha_{lj} + p_ip_l\Em_{\{i,l\}}\alpha_{il}\alpha_{li} -S_i -S_l +1 \nonumber\\
     &\hspace{1cm}- p_ip_lp_{il}p_{li}\alpha_{il}\alpha_{li} + S_iS_l  - \sum_{j \in [n]}p_j^2p_{ij}p_{lj}\alpha_{ij}\alpha_{lj}.
\end{align}
}

Hence, the second term of \eqref{eq:MSE_expression} is given by

{\small
\begin{align}
    &\frac{1}{n^2}\sum_{\substack{i,l \in [n] \\ i \neq l}}\left[\sum_{j \in [n]}p_j(1 - p_j)p_{ij}p_{lj}\alpha_{ij}\alpha_{lj}\right. \nonumber\\
    &\left. + p_ip_l(\Em_{\{i,l\}} - p_{il}p_{li})\alpha_{il}\alpha_{li} + S_iS_l -S_i -S_l + 1\right.\Biggr]\xv_i^{\top}\xv_l.
\end{align}
}

Subsequently, \eqref{eq:MSE_expression} simplifies to
{\small
\begin{align}
\label{eq:tiv_expression_second_term} 
    &\frac{1}{n^2}\sum_{i,j \in [n]}p_jp_{ij}\Paren{1 - p_{ij}}\alpha_{ij}^2\lVert \xv_i \rVert^2 + \frac{1}{n^2}\sum_{i \in [n]}(S_i - 1)^2\lVert \xv_i \rVert^2 \nonumber\\
    &\hspace{2cm}+ \frac{1}{n^2}\sum_{i,j,l \in [n]}p_j(1 - p_j)p_{ij}p_{lj}\alpha_{ij}\alpha_{lj}\xv_i^{\top}\xv_l \nonumber\\
    &\hspace{2cm}+\frac{1}{n^2}\sum_{i,l \in [n]}p_ip_l(\Em_{\{i,l\}} - p_{il}p_{li})\alpha_{il}\alpha_{li}\xv_i^{\top}\xv_l \nonumber\\
    &\hspace{2cm}+ \frac{1}{n^2} \sum_{\substack{i,l \in [n] \\ i \neq l}}(S_i - 1)(S_l - 1)\xv_i^{\top}\xv_l.
\end{align}
}

Here, the $i \neq l$ under the fourth summation vanishes because $\Em_{\{i,i\}} - p_{ii}p_{ii} = 1 - 1 \cdot 1 = 0$.
The above expression can be simplified as
{\small
\begin{align}
    &\frac{1}{n^2}\sum_{i,j \in [n]}p_jp_{ij}\Paren{1 - p_{ij}}\alpha_{ij}^2\lVert \xv_i \rVert^2 \nonumber\\
    &\hspace{10mm}+ \frac{1}{n^2}\sum_{i,j,l \in [n]}p_j(1 - p_j)p_{ij}p_{lj}\alpha_{ij}\alpha_{lj}\xv_i^{\top}\xv_l \nonumber\\
    &\hspace{10mm} +\frac{1}{n^2}\sum_{i,j \in [n]}p_ip_l(\Em_{\{i,l\}} - p_{il}p_{li})\alpha_{il}\alpha_{li}\xv_i^{\top}\xv_l \nonumber\\
    &\hspace{10mm} + \frac{1}{n^2} \sum_{\substack{i,l \in [n]}}(S_i - 1)(S_l - 1)\xv_i^{\top}\xv_j.
\end{align}
}

Recalling that ${\norm{\xv_i} \leq {\rm R}}$ for all $i \in [n]$, and using Cauchy-Schwarz inequality, we have $\lvert\xv_i^{\top}\xv_j\rvert \leq \norm{\xv_i}\norm{\xv_j} \leq {\rm R}^2$.
This yields an upper bound to the TIV given by
{\small
\begin{align*}
    &\sigma_{\rm tiv}^2(\pv, \Pv, \Av) = \frac{{\rm R}^2}{n^2}\left[\sum_{i,j \in [n]}p_jp_{ij}\Paren{1 - p_{ij}}\alpha_{ij}^2 \right. \nonumber\\
    &\hspace{15mm} \left.+ \sum_{i,j,l \in [n]}p_j(1 - p_j)p_{ij}p_{lj}\alpha_{ij}\alpha_{lj}\right. \\
    &\hspace{5mm}\left.+ \sum_{i,j \in [n]}p_ip_l(\Em_{\{i,l\}} - p_{il}p_{li})\alpha_{il}\alpha_{li} + \sum_{\substack{i,l \in [n]}}(S_i - 1)(S_l - 1)\right].
\end{align*}
}

Note that the last term can be simplified as
{\small
\begin{align}
    \sum_{i \in [n]} (S_i - 1)\sum_{l \in [n]} (S_l -1) = \Br{\sum_{i \in [n]}(S_i - 1)}^2.
\end{align}
}

So, we obtain the expression for TIV as in \eqref{eq:topology-induced-variance-expression}.

The last term is the bias in the global estimate at the PS.
For the unbiased case, we choose $\{\alpha_{ij}\}$ such that $S_i = 1$ for all $i$.
Consequently, the last term will disappear and we get the same upper bound for TIV as \cite{saha2022colrel, saha2023isit-p2p-privacy}.

The expression for {\rm PIV} is upper bounded similar to \cite{saha2023isit-p2p-privacy} but we present it here for completeness.
To simplify the {\rm PIV}, which depends on the privacy noise variance, we push $\tau_j$ inside $\sum_{i \in [n]}$ and interchange $i,j \in [n]$ to get
{\small
\begin{align}
    \label{eq:PIV_simplification}
    &\frac{1}{n^2}\mathbb{E}\left\lVert \sum_{i,j \in [n]}\tau_j \tau_{ij}\nv_{ij}\right\rVert_2^2 = \frac{1}{n^2}\sum_{i \in [n]}\mathbb{E}\left[\left\lVert \sum_{j \in [n]} \tau_j\tau_{ij}\nv_{ij} \right\rVert_2^2\right] \nonumber\\
    &\hspace{15mm}+\frac{1}{n^2}\sum_{\substack{i,l \in [n] \\ i \neq l}}\mathbb{E}\Br{\Paren{\sum_{j \in [n]}\tau_j\tau_{ij}\nv_{ij}}^{\hspace{-2mm}\top}\hspace{-1mm}\Paren{\sum_{m \in [n]}\tau_m\tau_{lm}\nv_{lm}}}.
\end{align}
}

Expanding the first term of \eqref{eq:PIV_simplification} yields
\begin{align}
    \underbrace{\frac{1}{n^2}\sum_{i,j \in [n]}p_jp_{ij}\sigma_{ij}^2d}_{:= \sigma_{\rm pr}^2(\pv, \Pv, \Sigmav)} + \underbrace{\frac{1}{n^2}\sum_{\substack{i,j,k \in [n] \\ j \neq k}}p_jp_kp_{ij}p_{ik}\mathbb{E}[\nv_{ij}^{\top}\nv_{ik}]}_{ = 0},
\end{align}
where the second term is zero due to our assumption of uncorrelated privacy noise, i.e., $\mathbb{E}[\nv_{ij}^{\top}\nv_{ik}] = 0$.
For the same reason, expanding the second term of \eqref{eq:PIV_simplification} yields
\begin{align}
    &\frac{1}{n^2}\sum_{\substack{i,l \in [n] \\ i \neq l}}\mathbb{E}\Br{\sum_{j, m \in [n]}\tau_j\tau_{ij}\tau_m\tau_{lm}\nv_{ij}^{\top}\nv_{lm}} \nonumber\\
    &\hspace{10mm}= \frac{1}{n^2}\sum_{\substack{i,l \in [n] \\ i \neq l}}\sum_{\substack{j, m \in [n],\\j\neq l, m\neq i}}p_jp_{ij}p_mp_{lm}\underbrace{\mathbb{E}\Br{\nv_{ij}^{\top}\nv_{lm}}}_{=0} \nonumber\\
    & \hspace{15mm}+ \frac{1}{n^2}\sum_{\substack{i,l \in [n] \\ i \neq l}}p_lp_iE_{\{i,l\}}\underbrace{\mathbb{E}\Br{\nv_{ij}^{\top}\nv_{ji}}}_{=0} = 0.
\end{align}

This completes the proof.

\vspace{-3mm}
\section{Privacy for protecting node identity at relay: Proof of Thm. \ref{thm:central_privacy_node_identity_protection}}
\label{app:privacy_node_identity_at_relay}

Note that when $i \notin \Rcal_j$, we have $\xtvu_j^{(-i)} = \xtvu_j$.
Consequently,
{\small
\begin{align}
    \label{eq:privacy_conditioning_on_node_participation}
    \Pr\left(\xtvu_j^{(-i)} \in {\cal S}\right) &= p_{ij}\Pr\left(\xtvu_j^{(-i)} \in {\cal S} \mid i \in \Rcal_j\right) \nonumber\\
    &\qquad + (1 - p_{ij})\Pr\left(\xtvu_j \in {\cal S} \mid i \notin \Rcal_j\right).
\end{align}
}
Let us denote the aggregated noise at node $j$ by $\nv_j$, where ${\nv_j \sim \Ncal(\mathbf{0}, \zeta_j^2\Iv_d)}$. 
The effective noise variance ${\zeta_j = \sum_{k \in \Rcal_j}\sigma_{kj}^2}$ is a random variable, with mean ${\zetao_j = \sum_{k \in [n]\setminus\{j\}}p_{kj}\sigma_{kj}^2}$.
Consider $\Ecal \triangleq \left\{\left\lvert \zeta_j - \zetao_j \right\rvert \leq r\right\}$, where $r$ is such that $\Pr(\Ecal^C) \leq \delta'$ for some $\delta' \in (0,1)$.
Then,
{\small
\begin{align}
\label{eq:eq:node_i_removed_probability_bound_1}
    \Pr\left(\xtvu_j^{(-i)} \in \Scal \mid i \in \Rcal_j\right) &= \Pr(\Ecal^C)\Pr\left(\xtvu_j^{(-i)} \in \Scal \mid i \in \Rcal_j, \Ecal^C\right) \nonumber\\
    &\quad\;\; + \Pr(\Ecal)\Pr\left(\xtvu_j^{(-i)} \in \Scal \mid 
     i \in \Rcal_j, \Ecal\right) \nonumber \\
    &\hspace{-15mm}\leq \delta' + \Pr(\Ecal)\Pr\left(\xtvu_j^{(-i)} \in \Scal \mid 
     i \in \Rcal_j, \Ecal\right).
\end{align}
}

Furthermore, ${\Pr\left(\xtvu_j^{(-i)} \in \Scal \mid i \in \Rcal_j, \Ecal\right)}$ can be written as
{\small
\begin{align}
\label{eq:node_i_removed_probability_bound_2}
     \sum_{\substack{\Acal \subseteq [n]\setminus\{j\} \\ i \in \Acal, \left\lvert \zeta_j - \zetao_j\right\rvert \leq r}} \hspace{-2mm}\Pr\left(\xtv_j^{(-i)} \in \Scal \mid \Rcal_j  = \Acal\right)\Pr\left(\Rcal_j = \Acal\right).
\end{align}
}

This allows us to analyze the privacy guarantee when the set of successfully transmitting nodes is fixed as $\Acal$.
For any $\zv \in \Real^d$, the {\it privacy loss} can be upper bounded as follows:
{\small
\begin{align}
\label{eq:privacy_loss_upper_bound_node_identity_protection}
    &\log\Paren{\frac{\Pr\Paren{\xtvu_j^{(-i)} = \zv \mid \Rcal_j = \Acal}}{\Pr\Paren{\xtvu_j = \zv \mid \Rcal_j = \Acal}}} \nonumber\\
    &= \log\Paren{\frac{\exp\Paren{-\frac{\left\lVert \zv - \sum_{k \in \Acal\setminus\{i\}}\alpha_{kj}\xv_k \right\rVert_2^2}{2\sum_{k \in \Acal\setminus\{i\}}\sigma_{kj}^2}}}{\exp\Paren{-\frac{\left\lVert \zv - \sum_{k \in \Acal}\alpha_{kj}\xv_k \right\rVert_2^2}{2\sum_{k \in \Acal}\sigma_{kj}^2}}}} \nonumber\\
    &\leq -\frac{1}{2\zeta_j}\Paren{\left\lVert \zv - \sum_{k \in \Acal\setminus\{i\}}\alpha_{kj}\xv_k \right\rVert_2^2 - \left\lVert \zv - \sum_{k \in \Acal}\alpha_{kj}\xv_k \right\rVert_2^2}
\end{align}
}

This upper bound is identical to the privacy loss of a Gaussian mechanism \cite{dwork2014algorithmic} with variance $\sum_{k \in \Acal}\sigma_{kj}^2$, with $\ell_2$ sensitivity given by
{\small
\begin{align}
    \sup_{\Acal : i \in \Acal} \left\lVert \sum_{k \in \Acal} \alpha_{kj}\xv_k - \sum_{k \in \Acal\setminus\{i\}}\alpha_{kj}\xv_k\right\rVert_2 = \lVert \alpha_{ij}\xv_i \rVert_2 \leq \alpha_{ij}\Rm,
\end{align}
}
where the last inequality follows since $\alpha_{ij} \geq 0$.
Consequently, from the guarantee of a Gaussian mechanism,
{\small
\begin{equation}
\label{eq:privacy_node_identity_fixed_participating_nodes}
    \Pr\Paren{\xtvu_j^{(-i)} \in \Scal \mid \Rcal_j = \Acal} \leq e^{\epsilon^{(p)}_{j \mid \Acal}} \Pr\Paren{\xtvu_j \in \Scal \mid \Rcal_j = \Acal} + \delta_j^{(p)},
\end{equation}
}
where,
{\small
\begin{equation}
\label{eq:privacy_node_identity_gaussian_mechanism_epsilon_fixed_participating_nodes}
    \epsilon^{(p)}_{j \mid \Acal} \triangleq \Br{2\log\Paren{\frac{1.25}{\delta_j^{(p)}}}}^{\frac{1}{2}}\frac{\alpha_{ij}\Rm}{\sqrt{\zeta_j}}.
\end{equation}
}
Let us denote $c = \Br{2\log\Paren{\frac{1.25}{\delta_j^{(p)}}}}^{\frac{1}{2}}\alpha_{ij}\Rm$.
From \eqref{eq:node_i_removed_probability_bound_2}, $\Pr\left(\xtvu_j^{(-i)} \in \Scal \mid i \in \Rcal_j, \Ecal\right)$ can be upper bounded by

{\small
\begin{align}
     &\sum_{\substack{\Acal \subseteq [n]\setminus\{j\} \\ i \in \Acal, \left\lvert \zeta_j - \zetao_j\right\rvert \leq r}}\hspace{-3mm}\Pr\Paren{\Rcal_j = \Acal}\Br{e^{\frac{c}{\sqrt{\zeta_j}}}\Pr\Paren{\xtvu_j \in \Scal \mid \Rcal_j = \Acal} + \delta_j^{(p)}} \nonumber\\
    &\leq e^{c\Paren{\zetao_j - r}^{-\frac{1}{2}}} \hspace{-6mm}\sum_{\substack{\Acal \subseteq [n]\setminus \{j\} \\ i \in \Acal, \left\lvert \zeta_j - \zetao_j\right\rvert \leq r}} \hspace{-5mm}\Pr\Paren{\Rcal_j = \Acal}\Pr\Paren{\xtvu_j \in \Scal \mid \Rcal_j = \Acal} + \delta_j^{(p)}.
\end{align}
}
The second inequality follows because $\zeta_j \geq \zetao_j - r$, and ${\sum_{\Acal \subseteq [n]\setminus\{j\}: i \in \Acal, \left\lvert \zeta_j - \zetao_j\right\rvert \leq r}\Pr(\Rcal_j = \Acal) \leq 1}$.
Using \eqref{eq:eq:node_i_removed_probability_bound_1}, we can upper bound $\Pr\Paren{\xtvu_j^{(-i)} \in \Scal \mid i \in \Rcal_j}$ by
{\small
\begin{align}
\label{eq:privacy_conditioning_on_node_participation_first_term}
     \delta' + e^{c\Paren{\zetao_j - r}^{-\frac{1}{2}}}\Pr\Paren{\xtvu_j \in \Scal \mid i \in \Rcal_j} + \delta_j^{(p)}.
\end{align}
}

Also, the second term of \eqref{eq:privacy_conditioning_on_node_participation} can be upper bounded by
{\small
\begin{align}
\label{eq:privacy_conditioning_on_node_participation_second_term}
    \Pr\Paren{\xtvu_j \in \Scal \mid i \notin \Rcal_j} \leq e^{c\Paren{\zetao_j - r}^{-\frac{1}{2}}}\Pr\Paren{\xtvu_j \in \Scal \mid i \notin \Rcal_j}.
\end{align}
}

From \eqref{eq:privacy_conditioning_on_node_participation_first_term} and \eqref{eq:privacy_conditioning_on_node_participation_second_term}, we upper bound $\Pr\Paren{\xtvu_j^{(-i)} \in \Scal}$ by
{\small
\begin{align}
\label{eq:privacy_node_identity_final_expression}
     &e^{c\Paren{\zetao_j - r}^{-\frac{1}{2}}}\left(p_{ij}\Pr\Paren{\xtvu_j \in \Scal \mid i \in \Rcal_j} \right.\nonumber\\
     &\qquad\left.+ (1 - p_{ij})\Pr\Paren{\xtvu_j \in \Scal \mid i \notin \Rcal_j}\right) + p_{ij}\Paren{\delta' + \delta_j^{(p)}} \nonumber\\
    &\leq e^{c\Paren{\zetao_j - r}^{-\frac{1}{2}}}\Pr\Paren{\xtvu_j \in \Scal} + p_{ij}\Paren{\delta' + \delta_j^{(p)}}.
\end{align}
}
This completes the proof.

\vspace{-5mm}
\section{Privacy for protecting local data at a relay: Proof of Thm. \ref{thm:central_privacy_node_dataset_protection}}
\label{app:privacy_local_data_at_relay}

The proof of Thm. \ref{thm:central_privacy_node_dataset_protection} is quite similar to that of Thm \ref{thm:central_privacy_node_identity_protection}, and we skip the details.
Similar to \eqref{eq:privacy_conditioning_on_node_participation}, since $\xtvu_j$ remains unchanged on perturbing $\xv_i$ if $i \notin \Rcal_j$, for any $\uv \in \Real^d$,
{\small
\begin{align}
    \label{eq:privacy_conditioning_on_local_data}
    &\Pr\Paren{\xtvu_j \in \Scal \mid \xv_i = \uv} = p_{ij}\Pr\Paren{\xtvu_j \in \Scal \mid \xv_i = \uv, i \in \Rcal_j} \nonumber\\
    &\qquad\qquad\qquad\;\; + (1 - p_{ij})\Pr\Paren{\xtvu_j \in \Scal \mid \xv_i = \uv', i \notin \Rcal_j}.
\end{align}
}
Conditioning on $\Ecal \triangleq \left\{\left\lvert \zeta_j - \zetao_j \right\rvert \leq r\right\}$ as in \eqref{eq:eq:node_i_removed_probability_bound_1}, $\Pr\left(\xtvu_j \in \Scal \mid \xv_i = \uv, i \in \Rcal_j\right)$ can be upper bounded by
{\small
\begin{align}
    \label{eq:eq:node_i_local_data_perturbed_probability_bound_1}
    \delta' + \Pr(\Ecal)\Pr\left(\xtvu_j \in \Scal \mid \xv_i = \uv', i \in \Rcal_j, \Ecal\right).
\end{align}
}
Moreover, similar to \eqref{eq:privacy_node_identity_fixed_participating_nodes}, for a fixed set of participating nodes $\Acal$, we also upper bound $\Pr\Paren{\xtvu_j \in \Scal \mid \xv_i = \uv, \Rcal_j = \Acal}$ by
\begin{equation}
     e^{\epsilon_{j \mid \Acal}^{(d)}}\Pr\Paren{\xtvu_j \in {\cal S} \mid \xv_i = \uv', \Rcal_j = \Acal} + \delta_j^{(d)}.
\end{equation}

Note that the $\ell_2$ sensitivity is now given by
{\small
\begin{equation*}
    \left\lVert \sum_{\substack{k \in \Acal \\ k \neq i}}\alpha_{kj}\xv_k + \alpha_{ij}\uv - \sum_{\substack{k \in \Acal \\ k \neq i}}\alpha_{kj}\xv_k - \alpha_{ij}\uv_i' \right\rVert_2 \leq 2\alpha_{ij}{\rm R}.
\end{equation*}
}
Here, we use $\lVert \uv - \uv' \rVert_2 \leq \lVert \uv \rVert + \lVert \uv' \rVert_2 \leq 2\Rm$. 
A tighter inequality can be used if we know an upper bound on the perturbation magnitude, i.e., $\lVert \uv - \uv' \rVert_2$.
As in \eqref{eq:privacy_node_identity_gaussian_mechanism_epsilon_fixed_participating_nodes}, we get,
{\small
\begin{align}
\label{eq:privacy_local_data_gaussian_mechanism_epsilon_fixed_participating_nodes}
    \epsilon^{(d)}_{j \mid \Acal} \triangleq \Br{2\log\Paren{\frac{1.25}{\delta_j^{(d)}}}}^{\frac{1}{2}}\frac{2\alpha_{ij}\Rm}{\sqrt{\zeta_j}}.
\end{align}
}

Similar to \eqref{eq:privacy_conditioning_on_node_participation_first_term}, denoting $c = \Br{2\log\Paren{\frac{1.25}{\delta_j^{(d)}}}}^{\frac{1}{2}}2\alpha_{ij}\Rm$, we upper bound $\Pr\left(\xtvu_j \in \Scal \mid \xv_i = \uv, i \in \Rcal_j\right)$ by
{\small
\begin{equation}
     \delta' + e^{\ct\Paren{\zetao_j - r}^{-\frac{1}{2}}}\Pr\Paren{\xtvu_j \in \Scal \mid \xv_i = \uv', i \in \Rcal_j} + \delta_j^{(d)}.
\end{equation}
}

Finally, similar to \eqref{eq:privacy_node_identity_final_expression}, we complete the proof by upper bounding $\Pr\Paren{\xtvu_j \in \Scal \mid \xv_i = \uv}$ by
{\small
\begin{equation}
    e^{\ct\Paren{\zetao_j - r}^{-\frac{1}{2}}}\Pr\Paren{\xtvu_j \in \Scal \mid \xv_i = \uv'} + p_{ij}\Paren{\delta' + \delta_j^{(d)}}.
\end{equation}
}

\section{Privacy when the PS can observe outputs of multiple relays: Proof of Thm. \ref{thm:central_privacy_at_PS}}
\label{app:proof_central_privacy_at_PS}

In Thm. \ref{thm:central_privacy_node_identity_protection}, let us denote $\delta'$ by $\delta_j'$ such that $\Pr\Paren{\left\lvert \zeta_j - \zetao_j\right\rvert \geq r_j} \leq \delta_j'$. 
When $p_{ij} > 0$, for any $\delta \in (0, p_{ij}]$, let us set $\delta_j^{(p)} = \frac{\delta}{p_{ij}} - \delta_j'$.
Then, for $\delta \in (0, p_{ij}]$, \eqref{eq:def_node_identity_protection_relay} is
{\small
\begin{align}
\label{eq:central_privacy_relay_restated}
    &\Pr\left(\xtv_j^{(-i)} \in \Scal\right) \leq \epsilont_{ij}^{(p)}\Pr\left(\xtv_j \in \Scal\right) + \delta, \quad \text{where} \nonumber\\ 
    &\epsilont_{ij}^{(p)} = \exp{\Paren{\Br{2\log\Paren{\frac{1.25\;p_{ij}}{\delta - p_{ij}\delta_j'}}}^{\frac{1}{2}}\frac{\alpha_{ij}\Rm}{\sqrt{\zetao_j + \sigma_{jj}^2 - r_j}}}}.
\end{align}
}

Here, $\sigma_{jj}^2$ appears in the denominator of $\epsilont_{ij}^{(p)}$ because we consider the signal transmitted by node $j$ to the PS, i.e., $\xtv_j$, instead of the signal aggregated at node $j$, i.e., $\xtvu_j$.
Since node $j$ adds $\nv_{jj} \sim \Ncal(\mathbf{0}, \sigma_{jj}\Iv_d)$ to privatize its own data before sending to the PS, this contributes to the effective privacy noise variance in $\xtv_j$.

Let us consider the privacy guarantee when the eavesdropper (possibly PS) can observe $\yv_j = \tau_j\xtv_j$.
Denote $\yv_j^{(-i)} = \tau_j\xtv_j^{(-i)}$.
Then, for any measurable set $\Scal$, $ \Pr(\yv_j^{(-i)} \in \Scal)$ is given by
{\small
\begin{align}
    &p_j\Pr\Paren{\yv_j^{(-i)} \in \Scal \mid \tau_j = 1} + (1 - p_j)\Pr\Paren{\yv_j^{(-i)} \in \Scal \mid \tau_j = 0} \nonumber\\
    &\stackrel{\rm (i)}{\leq} p_j\Paren{\epsilont_{ij}^{(p)}\Pr\left(\yv_j \in \Scal \mid \tau_j = 1\right) + \delta} \nonumber\\
    &\qquad\qquad\qquad\qquad+ (1 - p_j)\epsilont_{ij}^{(p)}\Pr\Paren{\yv_j \in \Scal \mid \tau_j = 0} \nonumber\\
    &= \epsilont_{ij}^{(p)}\Pr\Paren{\yv_j \in \Scal} + p_j\delta,
\end{align}
}
where $\delta \in (0,p_{ij}]$, and $\rm (i)$ follows from \eqref{eq:central_privacy_relay_restated}.

Note that when the PS observes the outputs of two relays, namely nodes $j$ and $\ell$, both $\yv_j$ and $\yv_{\ell}$ contain some component of $\xv_i$ due to the redundancy introduced as a consequence of collaboration.
The accumulated noise in $\yv_j$ comes from the following sources of stochasticity: $\rm (i)$ the local privacy noise added by node $i$ for transmission to node $j$, i.e., $\nv_{ij}$, $\rm (ii)$ the intermittent nature of the link $i \to j$, i.e., $\tau_{ij}$, and $\rm (iii)$ the intermittent nature of the link from node $j$ to the PS, i.e., $\tau_j$.
To consider the privacy leakage about the participation of node $i$, note that the corresponding quantities for node $\ell$, i.e., $\nv_{i\ell}$, $\tau_{i\ell}$, and $\tau_\ell$ are independent of those for node $j$.
An application of the standard composition theorem \ref{defn:basic_composition_DP} tells us that if the PS observes the tuple $(\yv_j, \yv_\ell)$, the resulting mechanism is $(\epsilont_{ij}^{(p)} + \epsilont_{i\ell}^{(p)}, \delta (p_j + p_\ell))$ differentially private, where $\delta \in (0, {\rm min}\{p_{ij}, p_{i\ell}\}]$.

With a similar argument, we can conclude that if the PS observes the outputs of multiple relays, i.e., $\{\yv_{j}\}_{j \in [n]}$, the resulting mechanism is $\left(\sum_{\substack{j \in [n] \\ p_{ij} > 0}}\epsilont_{ij}^{(p)}, \delta\Paren{\sum_{j \in [n]}p_j} \right)$ differentially private with respect to protecting the identity of node $i$, where $\delta \in (0, {\rm min}_{\substack{j \in [n] \\ p_{ij} > 0}}\{p_{ij}\}]$.
Here, we consider only the nodes $j$ such that $p_{ij} > 0$, because when $p_{ij} = 0$, the corresponding mechanism are $(0,0)$-differentially private, since there is no transmission.
This gives us \eqref{eq:central_privacy_PS_node_identity}.

Proceeding in a similar fashion, we can obtain the privacy leakage for protecting the local data of a node, i.e., \eqref{eq:central_privacy_PS_local_data}.

\section{Weight and noise variance optimization using projected gradient descent}
\label{app:weight-optimization-projected-gradient-descent}

Recall our notation $\rho_{ij} = \frac{2\Rm}{\underline{\epsilon}_{ij}}\Br{2\log\Paren{\frac{1.25}{\underline{\delta}_{ij}}}}^{\frac{1}{2}}$.
Referring to Fig. \ref{fig:weight-optimization-cone-constraints}, let us consider a point ${\rm P} \equiv (\alpha, \sigma)$ and suppose its projection onto the cone $\Kcal$ is denoted by ${\rm \Phat} \equiv (\alpha_{\Kcal}, \sigma_{\Kcal})$.
Any point ${\rm P}$ can lie in one of the following three regions:

\begin{figure}[h!]
\captionsetup{width=0.5\textwidth}
    \centering
    \vspace{-5mm}
    \includegraphics[width=0.4\linewidth]{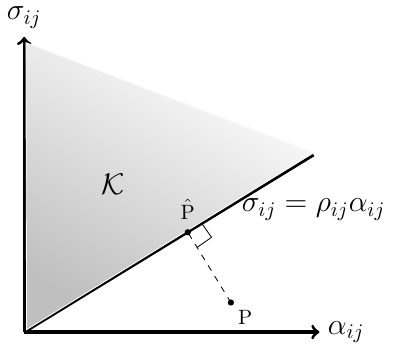}
    \caption{\small Projecting onto cone $\Kcal$ for finding $\alpha_{ij}$ and $\sigma_{ij}$.}
    \label{fig:weight-optimization-cone-constraints}
\end{figure}

{\bf Case $\mathbf{1}$}: $\alpha \geq 0$ and $\sigma \geq \rho_{ij}\alpha$. In this case, $(\alpha, \sigma) \in \Kcal$, and so, $(\alpha_{\Kcal}, \sigma_{\Kcal}) = (\alpha, \sigma)$.

{\bf Case $\mathbf{2}$}: $\sigma \geq 0$ and $\alpha < 0$.
The projection ${\rm \Phat} \equiv (\alpha_\Kcal, \sigma_\Kcal)$ is
{\small
\begin{equation}
    \alpha_{\Kcal} = 0 \; \text{ and }\; \sigma_{\Kcal} = \sigma.
\end{equation}
}

{\bf Case $\mathbf{3}$: } $\sigma < 0$ or $\{\sigma \geq 0 \; \text{ and }\; \sigma \leq \rho_{ij} \alpha\}$.
Let $(x,y)$ be the projection of the point on the line $\sigma = \rho_{ij}\alpha$.
Then, $(x,y)$ must satisfy $y = \rho_{ij}x$ and $(\alpha - x) +\rho_{ij} (\sigma - y) = 0$, i.e.,
{\small
\begin{equation}
x  = \frac{\alpha + \rho_{ij}\sigma}{1 + \rho_{ij}^2}, \text{ and } \; y =  \frac{\rho_{ij}(\alpha + \rho_{ij}\sigma)}{1 + \rho_{ij}^2}.
\end{equation}
}

Now, if $x < 0$, then the projection onto $\Kcal$ will be $(\alpha_{\Kcal}, \sigma_{\Kcal}) = (0, 0)$; otherwise leave them as it is.
This can be concisely expressed as
{\small
\begin{equation}
    \alpha_\Kcal = \Paren{\frac{\alpha + \rho_{ij}\sigma}{1 + \rho_{ij}^2}}_+ \; \text{ and } \; \sigma_\Kcal = \Paren{\frac{\rho_{ij}(\alpha + \rho_{ij}\sigma)}{1 + \rho_{ij}^2}}_+.
\end{equation}
}

Here, $\Paren{\cdot}_+ = {\rm max}\{\cdot, 0\}$ .
For each ${i,j \in [n]}$, we start with a feasible iterate $\Paren{\alpha_{ij}^{(0)}, \sigma_{ij}^{(0)}}$ that satisfies ${\alpha_{ij}^{(0)} \geq 0}$, ${\sigma_{ij}^{(0)} \geq \rho_{ij}\alpha_{ij}^{(0)}}$, and take projected gradient descent steps.

\section{Collaboration over Erdös-Rényi network topologies}
\label{app:collaboration-over-erdos-renyi-topologies}

{\bf Optimal collaboration weights}: For the setup described in \S \ref{subsec:erdos-renyi-collaboration}, the first term of the TIV \eqref{eq:topology-induced-variance-expression} is
{\small
\begin{align}
    &\frac{\Rm^2}{n^2}\sum_{i,j} p_jp_{ij}(1 - p_{ij})\alpha_{ij}^2 = \frac{\Rm^2}{n^2}\sum_{j \in \Mcal}q\sum_{i}p_{ij}(1 - p_{ij})\alpha_{ij}^2 \nonumber \\
    = &\frac{\Rm^2q}{n^2}\sum_{j \in \Mcal}\left[p_{jj}(1 - p_{jj})\alpha_{jj}^2 + \sum_{i \in \Mcal \setminus \{j\}}p_{ij}(1 - p_{ij})\alpha_{ij}^2\right. \nonumber\\
    & \quad\left. + \sum_{i \in \Mcalo}p_{ij}(1 - p_{ij})\alpha_{ij}^2\right] = \frac{\Rm^2}{n^2}m(n-m)qp(1-p)\alpha^2.
\end{align}
}
The second term of \eqref{eq:topology-induced-variance-expression}, $ {\frac{\Rm^2}{n^2}\sum_{i,j,l} p_j(1 - p_j)p_{ij}p_{lj}\alpha_{ij}\alpha_{lj}}$, is
{\small
\begin{align}
    &\frac{\Rm^2}{n^2}q(1-q)\sum_{j \in \Mcal}\sum_{i, l}p_{ij}p_{lj}\alpha_{ij}\alpha_{lj} \nonumber\\
    &= \frac{\Rm^2}{n^2}q(1-q)\sum_{j \in \Mcal}\left[\sum_{i}p_{ij}^2\alpha_{ij}^2 + \sum_{i,l : i \neq l}p_{ij}p_{lj}\alpha_{ij}\alpha_{lj}\right] \nonumber \\
    &= \frac{\Rm^2}{n^2}q(1-q)\sum_{j \in \Mcal}\left[\sum_{j \in \Mcal\setminus\{i\}}p_{ij}^2\alpha_{ij}^2 + p_{ii}^2\alpha_{ii}^2 + \sum_{j \in \Mcalo}p_{ij}^2\alpha_{ij}^2 \right.\nonumber\\
    & + \left. \sum_{l \in \Mcalo}p_{jj}p_{lj}\alpha_{jj}\alpha_{lj} + \sum_{i \in \Mcalo}p_{ij}p_{jj}\alpha_{ij}\alpha_{jj} + \sum_{i, l \in \Mcalo: i \neq l}\hspace{-3mm}p_{ij}p_{lj}\alpha_{ij}\alpha_{lj}\right] \nonumber \\
    &= \frac{\Rm^2}{n^2}q(1-q)\sum_{j \in \Mcal}\left[\gamma^2 + (n-m)p^2\alpha^2 + 2(n-m)p\gamma\alpha \right. \nonumber\\
    &\left.\qquad\qquad\qquad\qquad+ (n-m)(n-m-1)p^2\alpha^2\right] \nonumber\\
    &= \frac{\Rm^2}{n^2}mq(1-q)\left(\gamma + (n-m)p\alpha\right)^2.
\end{align}
}

The third term of \eqref{eq:topology-induced-variance-expression}, ${\frac{\Rm^2}{n^2}\sum_{i,j}p_ip_j(\Em_{ij} - p_{ij}p_{ji})\alpha_{ij}\alpha_{ji}}$, is
\begin{align}
     \frac{\Rm^2}{n^2}\sum_{i \in \Mcal}p_i^2(\Em_{ii} - p_{ii}^2)\alpha_{ii}^2 = 0.
\end{align}
This follows as $\alpha_{ij} = 0$ whenever $i, j \in \Mcal$, $i \neq j$, and because $\Em_{ii} = p_{ii} = 1$.

The fourth term of \eqref{eq:topology-induced-variance-expression}, ${\frac{\Rm^2}{n^2}\left(\sum_{i,j}p_jp_{ij}\alpha_{ij} - n\right)^2}$, is
{\small
\begin{align}
&\frac{\Rm^2}{n^2}\left(\sum_{j \in \Mcal}p_j\left(p_{jj}\alpha_{jj} + \sum_{i \in \Mcalo}p_{ij}\alpha_{ij}\right)\right)^2 \nonumber\\
&\qquad\qquad= \frac{\Rm^2}{n^2}\left[mq\left(\gamma + (n-m)p\alpha\right) - n\right]^2.
\end{align}
}

The PIV given by \eqref{eq:privacy-induced-variance-expression} simplifies to
{\small
\begin{align}
    \frac{d}{n^2}\sum_{j \in \Mcal}p_j\sum_{i \in \Mcalo}p_{ij}\sigma_{ij}^2 = \frac{d}{n^2}qpm(n-m)\sigma^2.
\end{align}
}
With $\ell_2$ regularization for bias, we have
{\small
\begin{align}
    \beta(\alpha, \gamma) &= \sum_{i \in [n]}\Paren{\sum_{j \in [n]}p_jp_{ij}\alpha_{ij} - 1}^2 \nonumber\\
    &= m\Paren{q\gamma - 1}^2 + (n-m)\Paren{mqp\alpha - 1}^2.
\end{align}
}

So, the optimization problem \eqref{eq:weight-optimization} is given by
{\small
\begin{align}
    &\argminimize_{\alpha, \gamma, \sigma} \sigma_{\rm tiv}^2(\alpha, \gamma) + \sigma_{\rm piv}^2(\sigma) + \lambda \beta(\alpha, \gamma) \nonumber\\
    &\quad \text{subject to} \quad \xi\Rm\frac{\alpha}{\sigma} \leq \epsilon, \alpha \geq 0, \sigma \geq 0.
\end{align}
}
Here, 
{\small
\begin{align}
    &\sigma_{\rm tiv}^2(\alpha, \gamma) = \frac{\Rm^2}{n^2}m(n-m)qp(1-p)\alpha^2 \nonumber\\
    &\qquad\qquad\qquad+ \frac{\Rm^2}{n^2}mq(1-q)\left(\gamma + (n-m)p\alpha\right)^2 \nonumber\\
    &\qquad\qquad\qquad+ \frac{\Rm^2}{n^2}\left[mq\left(\gamma + (n-m)p\alpha\right) - n\right]^2, \nonumber\\
    &\sigma_{\rm pr}^2(\sigma) = \frac{d}{n^2}qpm(n-m)\sigma^2, \nonumber\\
    &\beta(\alpha, \gamma) = m\Paren{q\gamma - 1}^2 + (n-m)\Paren{mqp\alpha - 1}^2, \nonumber\\
    \text{and} \quad& \xi = 2\Br{2\log\Paren{\frac{1.25}{\delta}}}^{\frac{1}{2}}.
\end{align} 
}
For $(\alpha, \gamma, \sigma)$ that satisfy $\sigma \geq \xi\Rm\frac{\alpha}{\epsilon}$, $\alpha \geq 0$, $\sigma \geq 0$, we have
{\small
\begin{align*}
    &{\rm min}_{\alpha, \gamma, \sigma} \; \sigma_{\rm tiv}^2(\alpha, \gamma) + \sigma_{\rm pr}^2(\sigma) + \lambda\beta(\alpha, \gamma) \nonumber\\
    \geq \;\; &{\rm min}_{\alpha, \gamma} \; \sigma_{\rm tiv}^2(\alpha, \gamma) + \frac{d}{n^2}qpm(n-m)\Paren{\xi\Rm\frac{\alpha}{\epsilon}}^2 + \lambda\beta(\alpha, \gamma),
\end{align*}
}
and the inequality is tight for $\sigma = \xi\Rm\frac{\alpha}{\epsilon}$.
Since the RHS is a smooth and convex function of $\alpha$ and $\gamma$, we can find the optimal values by finding the stationary points. 
Differentiating with respect to $\alpha$ and $\gamma$, and some algebra yields the optimal collaboration weights given in \eqref{eq:optimal_weights_erdos-renyi}.

{\bf Central privacy guarantee}: From Thm. \ref{thm:central_privacy_node_identity_protection}, ${p_{ij} = p > 0}$, ${\epsilon_{ij}^{(p)} = \Br{2\log\Paren{\frac{1.25}{\delta_{j}^{(p)}}}}^{\frac{1}{2}}\frac{\alpha_{ij}{\rm R}}{\sqrt{\zetao_j - r}}}$, where ${j \in \Mcal}$ acts as a relay, and ${i \in \Mcalo}$ is a node with no direct connectivity to the PS.
Fix ${\delta_{ij}^{(p)} = \delta}$ to be the same $\delta$ as the peer-to-peer privacy constraint.
Then, ${\epsilon_{ij}^{(p)} = \frac{\xi\Rm}{2mpq}\frac{1}{\sqrt{\zetao_j - r}}}$, where
\[{\zetao_j = \sum_{k \in \Mcalo}p_{kj}\sigma_{kj}^2 = (n - m)p\sigma_*^2},\]
and,
\[r = \Paren{\frac{\sigma_*^2}{6} + \frac{1}{2}\Paren{\frac{\sigma_*^2}{9} + \frac{4(n-m)p(1-p)\sigma_*^4}{\log(2/\delta')}}^{\frac{1}{2}}}\log\Paren{\frac{2}{\delta'}}.\]
Choose $\delta'$ such that $12(n-m)p(1-p)\sigma_*^2 = \log\Paren{2/\delta'}$.
Since $\delta' \in (0,1)$, and $\sigma_* = \frac{\xi\Rm}{mpq\epsilon}$, when $n > m$, the above choice is valid if
{\small
\begin{align}
    &\sigma_* \geq \Paren{\frac{\log 2}{12(n-m)p(1-p)}}^{\frac{1}{2}} \nonumber\\
    &\qquad\qquad\qquad\quad  \text{or, } \quad \epsilon \leq \frac{\xi\Rm}{mq}\Paren{\Paren{\frac{1-p}{p}}\frac{12(n-m)}{\log 2}}^{\frac{1}{2}}.
\end{align}
}

With the above constraint on $\epsilon$ and this choice of $\delta'$, we get
{\small
\begin{equation}
    \zetao_j - r = (n-m)p\sigma_*^2\Paren{1 - 2\sigma_*(\sigma_* + 2)(1-p)}.
\end{equation}
}
Since $(\zetao_j - r) > 0$, when $p > \frac{7}{8}$, it suffices to ensure
{\small
\begin{align}
    &\sigma_*(\sigma_* + 2) \leq (\sigma_* + 2)^2 \leq \frac{1}{2(1-p)} \quad \text{or, } \quad \sigma_* \leq \frac{1}{\sqrt{2(1-p)}}-2 \nonumber\\
    &\qquad\qquad\qquad\qquad\quad \text{or, } \quad \epsilon \geq \frac{\xi\Rm}{mpq}\Paren{\sqrt{\frac{1}{2(1-p)}}-2}.
\end{align}
}
So, the privacy leakage is
{\small
\begin{align}
    \epsilon_{ij}^{(p)} &= \frac{\xi\Rm}{2mpq}\Paren{(n-m)p\sigma_*^2\Paren{1 - 2\sigma_*(\sigma_* + 2)(1-p)}}^{-\frac{1}{2}} \nonumber\\
    &\sim \Ocal\Paren{\frac{1}{m\sqrt{n-m}}}.
\end{align}
}

\clearpage


\section{Preliminaries}
\label{app:preliminaries}

\begin{definition} 
{\bf ($(\epsilon_{i}, \delta_{i})$-Local Differential Privacy (LDP))} Let $\mathcal{D}_i$ be a set of all possible data points at node $i$. For node $i$, a randomized mechanism $\mathcal{M}_i: \mathcal{D}_{i} \rightarrow \Real^{d}$ is $(\epsilon_{i}, \delta_{i})$ {\it locally differentially private (LDP)} if for any $x,x' \in \mathcal{D}_i$, and any measurable subset $\mathcal{S} \subseteq \text{Range}(\mathcal{M}_i)$, we have
\begin{align}
    \operatorname{Pr}(\mathcal{M}_i(x) \in \mathcal{S}) \leq e^{\epsilon_{i}}  \operatorname{Pr}(\mathcal{M}_i(x') \in \mathcal{S}) + \delta_{i}.
\end{align}
\end{definition}

The setting when $\delta_{i} = 0$ is referred as pure $\epsilon_{i}$-LDP.
In this work, we analyze the privacy level achieved by the \pricer~ algorithm, which exploits the intermittent connectivity and adds synthetic noise perturbations to privatize its local data. 
We focus on analyzing the privacy leakage under an additive noise mechanism that is drawn from the Gaussian distribution. 
This well-known perturbation technique is known as the {\it Gaussian mechanism}, and it provides rigorous privacy guarantees, defined next. 

\begin{definition}
    \label{defn:Gaussian_mechanism}
    {\bf (Gaussian Mechanism \cite{dwork2014algorithmic})}
    Suppose a node wants to release a function $f(X)$ of an input $X$ subject to $(\epsilon, \delta)$-LDP. The Gaussian release mechanism is defined as
    {\small
    \begin{align}
    \mathcal{M}(X) \triangleq f(X) + \mathcal{N}(0, \sigma^{2} \mathbf{I}).
    \end{align}
    }
    If the $\ell_2$-sensitivity of the function is bounded by $\Delta_f$, i.e., $\| f(x) - f(x')\|_{2}\leq \Delta_f$, $\forall x, x'$, then for any $\delta \in (0,1]$, the Gaussian mechanism satisfies $(\epsilon, \delta)$-LDP, where 
    {\small
    \begin{align}
        \epsilon = \frac{\Delta_{f}}{\sigma} \sqrt{2 \log \frac{1.25}{\delta}}. \label{gaussian_noise_add}
    \end{align}
    }
\end{definition} 

\begin{definition} {\bf (Basic Composition Thm. of DP \cite{dwork2014algorithmic})} \label{defn:basic_composition_DP} If a randomized mechanism $\mathcal{M}_{i}$ satisfies $(\epsilon_{i}, \delta_{i})$-LDP for $i = 1, 2, \cdots, n$, then for any $\delta_{i} \in (0,1]$, their composition $\mathcal{M}$ defined by $\mathcal{M} \triangleq (\mathcal{M}_{1}, \mathcal{M}_{2}, \cdots, \mathcal{M}_{K})$ satisfies $(\sum_{i=1}^{n}\epsilon_{i}, \sum_{i=1}^{n} \delta_{i})$-LDP.
\end{definition}

\subsection{Bernstein's inequality}
\label{subapp:bernstein_inequality}

\begin{lemma} (Bernstein's Inequality \cite{mitzenmacher2017probability}) 
\label{lem:bernstein_inequality}
Let $\{X_k\}_{k=1}^{n}$ be a collection of zero-mean independent  random variables, where each $X_k$ is bounded by $M$, almost surely. Then, for any $r \geq 0$, we have 
{\small
\begin{align}
    \operatorname{Pr} \bigg(\sum_{i=1}^{n} X_k > r\bigg) \leq \exp \left[-\frac{r^{2}}{\sum_{k=1}^{n} E(X_k^2) + \frac{Mr}{3}}\right].
\end{align}
}
\end{lemma}

{\bf Choice of the parameter $r$ for the result in Cor. \ref{cor:choice_of_t_bernstein_inequality}}: 
Consider the random variables to be ${X_k = \Paren{\tau_{kj} - p_{kj}}\sigma_{kj}^2}$.
Then, $\Ebb[X_k] = 0$, $M = \max_{k \in [n]}\sigma_{kj}^2$, and ${\Ebb[X_k^2] = p_{kj}(1 - p_{kj})\sigma_{kj}^4}$.
An application of Bernstein inequality gives
{\small
\begin{align*}
    &\Pr(\lvert \zeta_j - \zetao_j \rvert \geq r) = \Pr\Paren{\left\lvert\sum_{k \in [n]\setminus\{j\}}(\tau_{kj} - p_{kj})\sigma_{kj}^2\right\rvert \geq r} \nonumber\\
    &\leq 2\exp\Br{-\frac{r^2}{\sum_{k \in [n]\setminus\{j\}}p_{kj}(1-p_{kj})\sigma_{kj}^4+ \frac{r\max_{k \in [n]\setminus\{j\}}\sigma_{kj}^2}{3}}}.
\end{align*}
}
For any $\delta' \in (0,1]$, this is a quadratic in $r$ and the right hand side expression can be tightened by choosing $r$ to be

\newpage
{\small
\begin{align}
    r &= \frac{\log(2/\delta')}{2}\left({\max_{k \in [n]\setminus\{j\}} \sigma_{kj}^{2}}/{3} \right. \nonumber\\
    &\left. + \sqrt{{\max_{k \in [n]\setminus\{j\}} \sigma_{kj}^{2}}/{9}  + \frac{4}{\log(2/\delta')} \left(\sum_{k \in [n]\setminus\{j\}}\hspace{-3mm} p_{kj} (1-p_{kj}) \sigma_{kj}^{4}\right)}\right).
\end{align}
}

\section{Local privacy guarantee for node-node communications: Proof of Thm. \ref{thm:local-privacy-node-node}}
\label{app:proof-local-privacy-node-node}

We begin by considering the cases of successful transmission, i.e., $\tau_{ij} = 1$ and unsuccessful transmission, i.e., $\tau_{ij} = 0$ separately.
When $\tau_{ij} = 0$, we have perfect privacy, i.e.,
{\small
\begin{align}
    \label{eq:no_transmission_perfect_privacy}
    \Pr\Paren{\xtv_{ij} \in \Scal \mid \xv_i \in {\cal D}_i, \tau_{ij} = 0} & \hspace{-1mm}=\hspace{-1mm} \Pr\Paren{\xtv_{ij} \in \Scal \mid \xv_i' \in {\cal D}_i, \tau_{ij} = 0},
\end{align}
}
since there is no transmission from node $i$, i.e., $\xtv_{ij} = \mathbf{0}$ irrespective of whether $\xv_i \in {\cal D}_i$ or $\xv_i' \in {\cal D}_i$.
When ${\tau_{ij} = 1}$, node $j$ receives ${\xtv_{ij} = \alpha_{ij}\xv_i + \nv_{ij}}$.
In this case, the $\ell_2$-sensitivity is
{\small
\begin{equation}
    \label{eq:l2_sensitivity_node_node}
    \sup_{\substack{\xv_i, \xv_i' \text{ s.t. } \\ \lVert \xv_i \rVert \leq {\rm R}, \lVert \xv_i' \rVert \leq {\rm R}}} \lVert \alpha_{ij}\xv_i - \alpha_{ij}\xv_i' \rVert \leq 2\alpha_{ij}{\rm R}.
\end{equation}
}
Using \eqref{eq:l2_sensitivity_node_node} above, and from the guarantees of the Gaussian mechanism \cite{dwork2014algorithmic}, when $\tau_{ij} = 1$, we have for any $\delta_{ij} \in (0,1)$, the transmission from node $i$ to node $j$ is $(\epsilon_{ij}, \delta_{ij})$ - differentially private, i.e.,
{\small
\begin{align}
    \label{eq:successful_transmission_privacy}
    &\Pr\Paren{\xtv_{ij} \in {\cal S} \mid \xv_i \in {\cal D}_i, \tau_{ij} = 1} \nonumber\\
    &\qquad\qquad\qquad\leq e^{\epsilon_{ij}}\Pr\Paren{\xtv_{ij} \in {\cal S} \mid \xv_i' \in {\cal D}_i, \tau_{ij} = 1} + \delta_{ij},
\end{align}
}
where
{\small
\begin{equation*}
    \epsilon_{ij} = \Br{2\log\Paren{\frac{1.25}{\delta_{ij}}}}^{\frac{1}{2}}\frac{2\alpha_{ij}{\rm R}}{\sigma_{ij}}.
\end{equation*}
}

Combining \eqref{eq:no_transmission_perfect_privacy} and \eqref{eq:successful_transmission_privacy}, we have
{\small
\begin{align}
    &\Pr\Paren{\xtv_{ij} \in {\cal S} \mid \xv_i \in {\cal D}_i} \nonumber \\
    &= p_{ij}\Pr\Paren{\xtv_{ij} \in {\cal S} \mid \xv_i \in {\cal D}_i, \tau_{ij} = 1} \nonumber\\
    &\qquad\qquad + (1 - p_{ij})\Pr\Paren{\xtv_{ij} \in {\cal S} \mid \xv_i \in {\cal D}_i, \tau_{ij} = 0} \nonumber \\
    &\leq p_{ij}\Paren{e^{\epsilon_{ij}}\Pr\Paren{\xtv_{ij} \in {\cal S} \mid \xv_i' \in {\cal D}_i, \tau_{ij} = 1} + \delta} \nonumber\\
    &\qquad\qquad + (1 - p_{ij})\Pr\Paren{\xtv_{ij} \in {\cal S} \mid \xv_i' \in {\cal D}_i, \tau_{ij} = 0} \nonumber \\
    &\stackrel{\rm (i)}{\leq} p_{ij}e^{\epsilon_{ij}}\Pr\Paren{\xtv_{ij} \in {\cal S} \mid \xv_i' \in {\cal D}_i, \tau_{ij} = 1} \nonumber\\
    &\qquad + (1 - p_{ij})e^{\epsilon_{ij}}\Pr\Paren{\xtv_{ij} \in {\cal S} \mid \xv_i' \in {\cal D}_i, \tau_{ij} = 0} + p_{ij}\delta_{ij} \nonumber \\
    &=e^{\epsilon_{ij}}\Pr\Paren{\xtv_{ij}\in{\cal S} \mid \xv_i' \in {\cal D}_i} + p_{ij}\delta_{ij}.
\end{align}
}
Here, $\rm (i)$ holds true since ${e^{\epsilon_{ij}} \geq 1}$.
Furthermore, note that when ${p_{ij} = 0}$, node $i$ can never transmit to node $j$ and as a consequence, we can choose ${\alpha_{ij} = 0}$.
This completes the proof.

\end{document}